\documentclass[12pt]{article}
\usepackage{amsmath,amsfonts}
\makeatletter \@addtoreset{equation}{section}
\renewcommand{\theequation}{\thesection.\arabic{equation}}
\usepackage{cite}
\usepackage{bm}
\usepackage{color}

\makeatother
\def\one{{\hbox{ 1\kern-.8mm l}}}
\newcommand{\Dslash}{\not{\hbox{\kern-4pt $D$}}}
\newcommand{\pdslash}{\not{\hbox{\kern-2pt $\partial$}}}
\newcommand{\be}{\begin{equation}}
\newcommand{\bea}{\begin{eqnarray}}
\newcommand{\eea}{\end{eqnarray}}
\newcommand{\ba}{\begin{array}}
\newcommand{\ea}{\end{array}}
\newcommand{\ee}{\end{equation}}

\newcommand{\BEQ}{\begin{equation}}     
\newcommand{\BEA}{\begin{eqnarray}}
\newcommand{\BD}{\begin{displaymath}}
\newcommand{\EEQ}{\end{equation}}       
\newcommand{\EEA}{\end{eqnarray}}
\newcommand{\ED}{\end{displaymath}}
\newcommand{\demi}{\frac{1}{2}}         
\newcommand{\wht}[1]{\widehat{#1}}      
\newcommand{\weps}{\wht{\,\epsilon\,}}

\renewcommand{\vec}[1]{\boldsymbol{#1}} 

\newcommand{\matz}[4] 
     {\mbox{${\begin{array}{cc} #1 & #2 \\ #3 & #4 \end{array}}$}}

\newcommand{\appsection}[2]{\setcounter{equation}{0}\setcounter{subsection}{0}
\section*{Appendix #1. #2}
\renewcommand{\theequation}{#1.\arabic{equation}}
              \renewcommand{\thesection}{#1} }

\begin{document}

\begin{titlepage}
\vspace*{1mm}%

\vspace*{15mm}%
\begin{center}

{{\Large {\bf Logarithmic Exotic Conformal Galilean Algebras}}}

\vspace*{15mm} \vspace*{1mm} { Malte Henkel$^{a}$, Ali Hosseiny$^{b,d}$, and Shahin Rouhani$^{c,d}$}

\vspace*{.4cm}

{\it ${}^a$ Groupe de Physique Statistique, 
Institut Jean Lamour (CNRS UMR 7198),\\ Universit\'e de Lorraine Nancy, \\B.P. 70239, F--54506 Vand{\oe}uvre-l\`es-Nancy Cedex, France.}

\vspace*{.4cm}

{\it ${}^b$ Department of Physics, Shahid Beheshti University \\
G.C., Evin, Tehran 19839, Iran. }

\vspace*{.4cm}

{\it ${}^c$ Department of Physics, Sharif University of Technology \\
P.O. Box 11165-9161, Tehran, Iran. }

\vspace*{.4cm}

{\it ${}^d$ School of Particles and Accelerators, \\ Institute for Research in Fundamental Sciences (IPM)\\
P.O. Box 19395-5531, Tehran, Iran. \\}

\vspace*{.4cm}

e-mail: { {\tt malte.henkel@univ-lorraine.fr},~~ \tt al$\_$hosseiny@sbu.ac.ir} ~~and {\tt rouhani@ipm.ir}

\vspace*{2cm}
\end{center}

\begin{abstract}

Logarithmic representations of the conformal Galilean algebra (CGA) 
and the Exotic Conformal Galilean algebra ({\sc ecga}) are constructed. 
This can be achieved by non-decomposable representations of the scaling dimensions or the rapidity indices, specific
to conformal galilean algebras. Logarithmic representations of the non-exotic CGA lead to the expected constraints on 
scaling dimensions and rapidities and also on the logarithmic contributions in the co-variant two-point functions. 
On the other hand, the {\sc ecga} admits several distinct situations which are distinguished by different sets of constraints
and distinct scaling forms of the two-point functions. Two distinct realisations for the spatial rotations are identified as well. 
The first example of a reducible, but non-decomposable representation, without logarithmic terms in the two-point function is given. 

\end{abstract}

\end{titlepage}



\section{Introduction}

Logarithmic conformal field theories (LCFT) arose, by noticing that the independent solutions 
of the null vector equation governing the behaviour of the four point function, could coincide in certain cases; 
giving rise to new independent solutions involving logarithms \cite{{Saleur},{Gurarie}}. 
Previously this possibility was ignored because unitarity ruled it out; 
however, applications for such non-unitary theories could be 
found within condensed-matter or statistical physics 
(for reviews of LCFT and applications see \cite{{Gaberdiel1},{Gaberdiel2},{Flohr},{LCFT},{Mathieu07}}). 
On another front, recent developments 
has attracted interest towards non-relativistic conformal field theories (NRCFT) 
\cite{{Niederer},{Hagen},{Henkel1994},{Duval1990},{Henkel02},{Son},{Balasubramanian},{AlishahihaRouhani},{Duval09},{Aizawa},{Henkel10}}. 
These are theories based on attempted extensions of the Galilean symmetries, 
the motivation being that they may apply to low-energy and/or 
time-dependent systems in condensed-matter or statistical physics. 
The best-known special cases of such symmetry algebras are the Schr\"odinger algebra and the 
{\em conformal Galilei algebra} (CGA), both to be defined below. 
The natural question arises as to whether logarithmic correlators may be found for such NRCFTs 
\cite{{Hosseinylog},{HosseinyNaseh},{Hotta}}, 
for a recent review see \cite{Rouhanihenkel}. The answer is affirmative. 
Furthermore, applications including the one-dimensional contact process (Reggeon field-theory) 
and the one-dimensional Kardar-Parisi-Zhang equation have been suggested \cite{{Henkellog},{Henkelkpz}}. 
In this paper, we present new
logarithmic correlators for the Exotic Galilean Algebra ({\sc ecga}) \cite{{Lukierski1},{Lukierski2}}, 
which is CGA  in 2+1 dimensions, but with an `exotic' central charge.

Naturally, non-relativistic conformal symmetries are based on Galilean symmetry. 
A Galilean transformation ($\vec{x}\mapsto \vec{x}', t\mapsto t'$) acts on a point $\vec{x}$ 
in $d$-dimensional Euclidean space, at a given time $t$, according to:
\bea
\vec{x}{'}=R\vec{x}+\vec{b}t+\vec{a}, \;\;\;\;\;\;\;\;\;\;\;t{'}=t+c
\eea
where $R\in\mbox{\sl SO}(d)$ is a $d\times d$ 
rotation matrix, $\vec{b}$ and $\vec{a}$ are $d$-dimensional vectors and $c$ is a constant. 
However, we shall look at larger symmetries. For instance the symmetry group 
(called the {\it Schr\"odinger group}) 
of the free Schr\"odinger equation is larger:
\BEQ \label{1.2}
\vec{x}{'}=\frac{R\vec{x}+\vec{b}t+\vec{a}}{ft+g}, \;\;\;\;\;\;\;\;\;t{'}=\frac{dt+c}{ft+g}\;\;\;\;\;\;\;\;\;\;dg-fc=1
\EEQ
The Lie algebra ({\it Schr\"odinger algebra}) spanned by the 
infinitesimal generators of the transformations (\ref{1.2}) 
is given below for $1+1$ dimensions. Being non-semi-simple, 
this Lie algebra admits a non-trivial central charge, related to
projective transformations of the solutions of the Schr\"odinger equation. 
It is related to the (non-relativistic) `mass' ${\cal M}$ 
of the system. This can be generalised straightforwardly to what we shall call 
{\it $l$-Galilei algebras}\footnote{In some papers these are referred to as 
$spin-l \; Galilei$ algebras \cite{Tachikawa}, 
however the index $l$ has nothing to do with spin.} 
by admitting a more complex transformation \cite{{Henkel1997},{Negro}};
\BEQ \label{1.3}
\vec{x}{'}=\frac{R\vec{x}+\vec{b}_{2l}t^{2l}+....+\vec{b}_1t+\vec{b}_0}{(ft+g)^{2l}}, 
\;\;\;\;\;\;\;\;\;t{'}=\frac{dt+c}{ft+g}\;\;\;\;\;\;\;\;\;\;dg-fc=1
\EEQ
where the $\vec{b}_i$, $i=0,1,\ldots,2l$ are $d$-dimensional vectors.
The transformations (\ref{1.3}) form a closed set and their infinitesimal generators span a closed Lie algebra
only for $l\in\demi \mathbb{Z}$ half-integer or integer. 
The Schr\"odinger group and its Lie algebra are recovered for $l=\demi$;
the case $l=1$ gives the {\em conformal Galilei group} 
and its Lie algebra, the {\em Conformal Galilean Algebra} (CGA) \cite{Havas}. 
These transformations, and more generally those of (\ref{1.3}), 
have in common the existence of a well-defined {\it dynamical exponent} $z$
such that under a dilatation
\BEQ
\vec{x}\rightarrow \lambda \vec{x} \;\; ,\;\; t \rightarrow \lambda^z t
\EEQ
such that $z=1/l$ for the $l$-Galilei transformations (\ref{1.3}). 
The two important special cases $l=\demi,1$ also arise from two distinct more general approaches:
\begin{enumerate}
\item One may try to generalise (\ref{1.3}) further by extending the projective conformal 
(or M\"obius) transformations in the time $t$
to arbitrary conformal transformations. 
Taking the projective terms describing the transformation of the wave functions into account, the
only cases with local generators which close as a Lie algebra are, 
besides evidently conformal transformations in space-time, 
the cases $l=\demi,1$ of the
Schr\"odinger algebra and the CGA \cite{Henkel02}.
\item When considering the non-relativistic limit of space-time conformal transformations 
and assuming the existence of a 
dynamical exponent $z$, restriction to time-like and 
light-like geodesics reproduces exactly the Schr\"odinger algebra and the CGA, for
$z=2$ and $z=1$, respectively \cite{Duval09}. 
\end{enumerate}
Physical applications either refer to strongly anisotropic systems at equilibrium, where the `time' 
$t$ is just a name for a peculiar spatial
direction with strongly modified interactions such that $z=\theta$ 
is better referred to as an `anisotropy exponent' 
(paradigmatic examples are uniaxial Lifshitz points in lattice spin models with competing interactions); 
or else to real dynamics, at or far away from equilibrium. 
In the first case, co-variant $n$-point functions 
(such as we shall calculate later on) will represent physical correlators; the second
case, causality constraints\footnote{An algebraic method of derivation uses an embedding 
into a parabolic sub-algebra of the
conformal algebra in $d+2$ dimensions, see \cite{causality} for details.} 
imply that $n$-point functions are to be interpreted as 
response functions with respect to some external
perturbation. See \cite{Henkel10} for an introduction and overview on recent results. 
For brevity, we shall refer throughout to the
two-point functions to be computed as `correlators'.  

Finally returning to the Lie algebra of the symmetry transformations (\ref{1.3}), 
in $1+1$ dimensions it spanned by the generators: 
\bea\label{1.5} 
\begin{split}
&H=-\partial_{t},\;\;\;\;\;\;\;\;\;\;\;\;\;\;\;\;\;\;\;\;\;\;\;\;\;\;\;\;\;\;
P^n=-t^n\partial_{x}\;\;;\;\;\;\;\;\;\;\;\;\;\;n=0,...,2l,\cr&
D=-(t\partial_{t}+lx\partial_{x}),\;\;\;\;\;\;\;\;\;\;\;\;\;\;\;
C=-(2ltx\partial_{x}+t^2\partial_{t}).
\end{split}
\eea
with the following non-vanishing commutators
\bea \label{1.6} 
\begin{split}
&[D,H]=H,\;\;\;\;\;\;\;\;\;\;\;\;\;\;\;\;\;\;\;\;\;\;\;\;\;\;[D,C]=-C, \;\;\;\;\;\;\;\;\;\;\;\;\;\;\;\;\;\;[C,H]=2D,
\cr& [D,P^n]=(l-n)P^n,\;\;\;\;\;\;\;\;\;\;\;\;\;[H,P^n]=-nP^{n-1},\;\;\;\;\;\;\;\;\;\;[C,P^n]=(2l-n) P^{n+1}.
\end{split}
\eea
Known physical realisations of these algebras are known for $l=1/2$ 
as the Schr\"odinger algebra\footnote{Especially 
in the phase-ordering kinetics far from equilibrium for spin systems quenched to temperatures 
$T<T_c$ below the critical temperature $T_c>0$, when for a non-conserved order parameter one has naturally 
$z=2$ \cite{{Bray94},{HenkelPleim},{Henkel10}}.}, for $l=1$ as the CGA and 
for $l=2$ and $l=3$ in the Lifshitz points of first and second order in the 
ANNNS model, which adds uniaxial competing interactions 
to the so-called spherical model.\footnote{See \cite{Henkel1997,Henkel02}. 
When considering the uniaxial Lifshitz points in systems with competing interactions such as the
ANNNO($n$) model, field-theoretic two-loop calculations have shown that the anisotropy exponent 
$\theta-\demi = \mbox{\rm O}(\varepsilon^2)$ in $d=4.5-\varepsilon$ 
dimensions or $\theta-\demi=\mbox{\rm O}(1/n)$, 
which known, non-vanishing coefficients which are of the order
$\approx 10^{-3}-10^{-2}$ \cite{{DiehlShpot00},{ShpotDiehl01},{ShpotDiehl05},{ShpotDiehl08}}. 
The ANNNS model corresponds to $n\to\infty$.} 
It remains an open problem to find physical realisations for generic values of $l$. 

CGA is special because it can be obtained from the relativistic conformal algebra through contraction. 
When contracting, in some sense we are investigating the symmetry for low velocities. 
In other words we allow:
\bea\label{contraction} 
\vec{x}\rightarrow \frac{\vec{x}}{c},\;\;\;\;\;\;\;\;\;\;\;\;\;\;\;\;\; t \rightarrow
t,\;\;\;\;\;\;\;\;\;\;\;\;\;\;\;\;\;\;c\rightarrow \infty. 
\eea
In $1+1$ dimensions, CGA is even more special since it has an infinite-dimensional extension (which is called 
`full CGA/altern-Virasoro algebra' in the literature, contains a Virasoro sub-algebra and admits
two independent central charges \cite{OvsienkoRoger98}) which in turn can be obtained fully from 
contraction \cite{{Henkel2006}, {HosseinyRouhani}}. This 
infinite-dimensional extension of the CGA is almost solvable \cite{Bagchi2d}, 
a property which helps to investigate logarithmic 
representations and holographic realisation easily \cite{{Hosseinylog}, {HosseinyNaseh}}.

Here, we study some properties of the finite-dimensional CGA 
(and leave aside its infinite-dimensional extensions). 
In $2+1$ dimensions, CGA admits a non-trivial central extension 
(the so-called ``exotic" central charge \cite{Lukierski1,Lukierski2}) which forbids 
Galilean boosts to commute, reminiscent of non-commutative theories. 
Its physical significance has been of interest 
\cite{DuvalHorvathy,Cherniha10}. The central charge can also be obtained 
by contraction and two-point function is realised using auxiliary 
coordinates \cite{Tachikawa}. In this paper we consider this exotic algebra 
{\sc ecga} and show that logarithmic representations exist. A new feature
arises in the CGA and the {\sc ecga} 
in that the rapidities allow for extra types of logarithmic representations, which we shall construct. 
We work out two-point functions for realisations in which the rapidity index is included. 
The `exotic' extension of the CGA in $2+1$ dimensions leads to several unexpected results on the
form of the two-point functions; notably, we discuss the consequences of two distinct realisations of
rotation-invariance (which from a purely algebraic point of view are indistinguishable). 
We hope these results to be useful in future attempts in identifying specific models with conformal galilean symmetries. 

This paper is organised as follows: In section 2 we give a very brief presentation of LCFT, 
and recall the derivation of the two-point functions in logarithmic representations of the LCFT, the 
Schr\"odinger algebra and the CGA, using the elegant formalism of nil-potent variables.  
In section 3 we give a short introduction to the exotic CGA, and derive the two-point functions, both for scalar and
logarithmic representations. 
Some conclusions are presented in section~4, with a table summarising our findings in a compact manner.  
Several appendices treat technical aspects of the calculations, either in the {\sc ecga} or on rotation-invariance.

\section{Logarithmic CFT: background}

\subsection{Basic formalism}

Logarithmic conformal field theories (LCFTs) arise when indecomposable 
but reducible representations of the Virasoro algebra are taken 
\cite{{Saleur},{Gurarie}}  (for reviews see \cite{{Gaberdiel1},{Gaberdiel2},{Flohr},{LCFT}}). 
When the action of the scaling operator on the Verma module is not diagonal it gives rise to staggered modules 
\cite{{KytolaRidout},{Ridout}}. In the simplest case, the highest weight primary operator and 
its logarithmic partner form a rank-$2$ Jordan cell:
\bea\label{fieldJC}
L_{0}\phi_h(Z)|0\rangle =h\phi_h(Z)|0\rangle ,\;\;\;\;\;\;\;\;\;\;\;\;\;\;\;\;\;\;\;
L_{0}\psi_h(Z)|0\rangle =h\psi_h(Z)|0\rangle +\phi_h(Z)|0\rangle.
\eea
There is a simple method for dealing with case by introducing nilpotent variables 
$\theta_i$ which satisfy the following relations:
\bea
\theta^2_i=0,\;\;\;\;\;\;\;\;\;\;\;\;\;\;\;\;\;\;\;\;\;\;\;\;\;\theta_i\theta_j=\theta_j\theta_i.
\eea
These nilpotent variables also admit complex conjugation which go into 
the anti-holomorphic part of the primary operators:
\bea
\bar\theta^2_i=0,\;\;\;\;\;\;\;\;\;\;\;\;\;\;\;\;\;\;\;\;\;\;\;\;\;\bar\theta_i\theta_j=\theta_j\bar\theta_i.
\eea
Now we can define our super-fields as
\bea\label{Superfield}
\Phi(z,\theta)=\phi(z)+\theta\psi(z),
\eea
and thereby equation ($\ref{fieldJC}$) is written compactly as \cite{RouhaniMoghimi}:
\bea\label{JordanEV}
L_{0}|h+\theta\rangle =(h+\theta)|h+\theta\rangle ,
\eea
where the state $|h+\theta\rangle$ is:
\bea
|h+\theta\rangle =|h,0\rangle+\theta|h,1\rangle.
\eea
This method allows a quick calculation of the two-point function. 
Concentrating on the holomorphic part of quasi-primary operators we obtain \cite{RouhaniMoghimi}:
\bea\label{twopoint1}
G(z_1,\theta_1;z_2,\theta_2 )=\langle\Phi_1(z_1,\theta_1)\Phi_2(z_2,\theta_2)\rangle
=g(\theta_1,\theta_2)(z_1-z_2)^{-(h_1+\theta_1+h_2+\theta_2)}\delta_{h_1,h_2}.
\eea
where $g(\theta_1,\theta_2)$ is given by
\bea
g(\theta_1,\theta_2)=a(\theta_1+\theta_2)+b\theta_1\theta_2.
\eea
and $a,b$ are normalisation constants. 
Now, expanding both sides of ($\ref{twopoint1}$) in powers of $\theta_{1,2}$,
one obtains the well-known logarithmic CFT two-point functions (with $z:=z_1-z_2)$)
\bea \label{eq:lconf} \begin{split}
&\langle\phi(z_1)\phi(z_2)\rangle=0,
\cr&\langle\phi(z_1)\psi(z_2)\rangle=a\,z^{-2h_1}\delta_{h_1,h_2}, \cr& 
\langle\psi(z_1)\psi(z_2)\rangle=z^{-2h_1}\,\left(b-2a\ln{z}\right)\delta_{h_1,h_2}.
\end{split}
\eea
This offers a simple and fast way of obtaining logarithmic correlators in other algebras as well, 
as we shall demonstrate below. 
Of course, we merely discussed here the most simple scenario for the appearance 
of logarithmic representations and  shall leave to future
work the description of more complex situations.


\subsection{Logarithmic representations of the Schr\"odinger algebra}

The Schr\"odinger algebra is the symmetry of the free Schr\"odinger equation. 
It is naturally tied in with Galilean symmetry. 
It is the smallest $(l=1/2)$ element of  the $\it{l-Galilei}$ algebras, plus a central extension:
\BEQ \label{2.10}
[P_i^0,P_j^1]={\cal M} \delta_{ij}
\EEQ
where the scalar $\cal M$ is the non-relativistic mass and $i,j=1,\ldots,d$. 
The Schr\"odinger algebra $\mathfrak{sch}(d)$ has a well-known infinite-dimensional extension
(with a Virasoro sub-algebra) which is now usually called the 
`Schr\"odinger-Virasoro algebra' ($\mathfrak {sv}$)\cite{Henkel1994}.  
In $1+1$ dimensions, the algebra $\mathfrak {sv}$ is represented by differential operators as 
(with $n\in\mathbb{Z}$ and $m\in\mathbb{Z}+\demi$):
\bea \label{2.11} \begin{split}
&X^n=-t^{n+1}\partial_t-\frac{1}{2}(n+1)t^nx\partial_x-\frac{1}{4}n(n+1){\cal{M}} t^{n-1}x^2-(n+1)ht^n,
\cr&Y^m=-t^{m+1/2}\partial_x-(m+\frac{1}{2})t^{m-1/2}{\cal{M}}x,
\cr&M^n=-{\cal{M}}t^n.
\end{split}
\eea
These generators make up the dynamical symmetry algebra of the free Schr\"odinger equation ${\cal S}\phi=0$, with 
the Schr\"odinger operator ${\cal S} := 2{\cal M}\partial_t -\partial_x^2 = 2M^0 X^{-1} -(Y^{-\demi})^{2}$. All generators (\ref{2.11})
of $\mathfrak{sch}(1):=\langle X^{0, \pm1}, Y^{\pm\frac{1}{2}},  M^0\rangle$ 
commute with $\cal S$, with the two  exceptions
\BEQ
{}\left[ {\cal S}, X^0 \right] = -{\cal S} \;\; , \;\; \left[ {\cal S}, X^1\right] = -2tS - 2{\cal M} \left( h-\frac{1}{2}\right)
\EEQ 
such that solutions of ${\cal S}\phi=0$ which have $h=\demi$ will be mapped onto another solution (and an obvious generalisation to
$d\geq 1$ dimensions).\footnote{There is an unitary bound
$h\geq d/2$ for the Schr\"odinger algebra $\mathfrak{sch}(d)$ in $d$ dimensions \cite{LeeLeeLee}.}
Representations of this algebra $\mathfrak{sch}(1)$ may be constructed by invoking scaling states:
\bea
X^0|h\rangle=0.
\eea
Now, similar to CFT, a rank 2 logarithmic representation may be found where two states exist, 
$|h,1\rangle $ and $|h,2\rangle$ such that the action of $X^0$ on them is non-diagonizable
\bea \label{2.14}
X^0|h,1\rangle=0,\;\;\;\;\;\;\;\;\;\;\;\;\;\;\;\;\;\;\;X^0|h,2\rangle=|h,1\rangle.
\eea
We follow the formalism of the previous sub-section.\footnote{At first sight, 
one might believe that because of the commutator 
(\ref{2.10}), with ${\cal M}\ne 0$, 
invariance under space-translations and Galilei-transformations could not be required simultaneously.
However, invariance under $M^0$ gives the Bargman super-selection rule 
${\cal M}^{[2]}={\cal M}_1 + {\cal M}_2=0$. Hence the action
of the commutator (\ref{2.10}) vanishes on any $n$-point function.} 
The two-point function is well known\footnote{Here, the `complex conjugate' 
$\Phi^*$ is obtained from $\Phi$ by changing the sign of the mass: 
${\cal M}\mapsto -{\cal M}$ \cite{Henkel10}.} \cite{Hosseinylog}:
\BEA
\langle\Phi_1(x_1, t_1, \theta_1), \Phi^*_2(x_2, t_2, \theta_2)\rangle &=& 
\delta_{h_1,h_2}\delta_{{\cal{M}}_1,{\cal{M}}_2} \: 
t^{-2h_1}\,
\exp\left[{-\frac{{\cal{M}}_1x^2}{2t}}\right] \nonumber \\
& & \times \left( b(\theta_1+\bar{\theta}_2)+\theta_1\bar{\theta}_2(c-2b\ln{t})\right)
\EEA
where $t:=t_1-t_2$, $x:=x_1-x_2$ and $b,c$ are normalisation constants. Expanding, one has 
\BEA
\langle\phi_1(t_1,x_1)\phi_2^*(t_2,x_2)\rangle &=& 0, \nonumber \\
\langle\phi(t_1,x_1)\psi_2^*(t_2,x_2)\rangle &=&
b\, t^{-2h_1}\,\exp\left[{-\frac{{\cal{M}}_1x^2}{2t}}\right]\,\delta_{h_1,h_2}\delta_{{\cal{M}}_1,{\cal{M}}_2}, \label{eq:lsch} \\ 
\langle\psi_1(t_1,x_1)\psi_2^*(t_2,x_2)\rangle &=& 
t^{-2h_1}\,\left(c-2b\ln{t}\right)\,\exp\left[{-\frac{{\cal{M}}_1x^2}{2t}}\right]\, 
\delta_{h_1,h_2}\delta_{{\cal{M}}_1,{\cal{M}}_2}.
\nonumber
\EEA


\subsection{Logarithmic CGA in $1+1$ dimensions}

Galilean conformal algebra in $1+1$ and $2+1$ dimensions is special. 
In $1+1$ dimensions, it is unique because it can be obtained directly from contracting 2-dimensional CFTs. 
Following this contraction many aspects of the fields can be extracted from CFT$_2$. 
In $2+1$ dimensions, it admits an `exotic' central charge \cite{{Lukierski1},{Lukierski2}}. 

For the moment, and to remain faithful to the method of previous section , consider 1+1 dimensions:
\bea \label{cga} \begin{split}
&P=-\partial_x\;,\;\;\;\;\;\;\;\;\;\;\;\;\;\;\;\;\;\;\;\;\;\;\;\;\;K=-t\partial_x-\gamma,\;\;\;\;\;\;\;\;\;\;\;\;\;\;\;\;\;\;\;\;\;\;\;\;\;\;\;F=-t^2\partial_x-2t\gamma,   
\cr& H =-\partial_t\;,\;\;\;\;\;\;\;\;\;\;\;\;\;\;D=-(t\partial_t+x\partial_x)-\Delta,\;\;\;\;\;\;\;\;\;\;\; C=-(2tx\partial_x+t^2\partial_t)-2t\Delta\;,
\end{split}\eea 
in which $\Delta$ is eigenvalue of dilation $D$ and $\gamma$ is eigenvalue of $K$ which is called $rapidity$. These can be further embedded into an infinite-dimensional 
set of generators (where $X^{-1,0,1}=H,D,C$ and $Y^{-1,0,1}=P,K,F$) which generate the infinite-dimensional Lie algebra
called `Full CGA/altern-Virasoro algebra' in the literature ($n\in\mathbb{Z}$):
\bea\label{gcageneratrs}
\begin{split}
&X^n=-\left[t^{n+1}\partial_t+(n+1)t^nx\partial_x+(n+1)(t^n\Delta+nt^{n-1}\gamma{x})\right]
\cr& Y^n=-\left[t^{n+1}\partial_x+(n+1)t^n\gamma\right]
\end{split}
\eea
with the commutators
\BEQ \label{2.18}
{}[X^m,X^n]=(m-n)X^{m+n} \;\; ,\;\; [X^m,Y^n]=(m-n)Y^{m+n} \;\; , \;\;  [Y^m,Y^n]=0.
\EEQ
Co-variant two-point functions are \cite{Henkel02,Henkel2006,BagchiRep} (with $x := x_1-x_2$ and $t :=t_1-t_2$)
\bea \label{eq:F_cga}
\langle\phi_1(t_1,x_1)\phi_2(t_2,x_2)\rangle_{\rm CGA}
=a\,\delta_{\Delta_1,\Delta_2}\delta_{\gamma_1,\gamma_2}\,t^{-2\Delta_1}\,
\exp\left[-{\frac{2\gamma_1 x}{t}}\right]
\eea
As mentioned above, the interesting point regarding CGA in $d=1+1$ dimensions 
is that we can obtain them from two-dimensional conformal symmetry 
by contraction. Briefly, $d=2$ conformal symmetry consists of two commuting Virasoro algebras, with generators:
\bea
L_{n}\;=\;-z^{n+1}\partial_{z},\;\;\;\;\;\;\;\;\;\;\overline{L}_{n}\;=\;-\overline{z}^{n+1}\partial_{\overline{z}}.
\eea
Under the contraction limit ($\ref{contraction}$), one observes that:
\bea\label{contractoperator}\begin{split}
&X^n=L^n+\overline{L}^n
\cr& Y^n=\frac{1}{c}(L^n-\overline{L}^n)
\end{split}\eea
generate the algebra given by (\ref{2.18}). 
The central charges of the two chiral copies of Virasoro algebra, 
namely $C$ and $\bar C$ combine to give the two independent 
central charges in the CGA, making it non-unitary \cite{Bagchi2d}. 
are a contracted limit of CFT$_2$,
it might be possible that its representations are contracted limit of CFT$_2$ representations \cite{Bagchi2d}. 
To observe this consider primary states in CFT$_2$:
\bea\label{eigenh}\begin{split}
&X^{0}|h,\overline{h}\rangle=(L_{0}+\overline{L}_{0})|h,\overline{h}\rangle
\;=\;(h+\overline{h})|h,\overline{h}\rangle,&\cr&
Y^{0}|h,\overline{h}\rangle\;=\;\frac{L_{0}-\overline{L}_{0}}{c}|h,\overline{h}\rangle\
\;=\;\frac{h-\overline{h}}{c}|h,\overline{h}\rangle.
\end{split}\eea
We observe that the scaling states of CFT$_2$ are scaling states of CGA, too. 
Now, they are identified by their scaling weight and rapidity. In other words
\bea
|h,\overline{h}\rangle\;\;\;\rightarrow\;\;\;|\Delta,\gamma\rangle,
\eea
in which
\bea\begin{split}
&\Delta:=h+\overline{h}
\cr&\gamma:=\frac{h-\overline{h}}{c}
\end{split}\eea

Now, to build a logarithmic representation of the full CGA, 
we expect logarithmic partners to appear in ($\ref{eigenh}$). 
The standard way to introduce them is to formally replace the real numbers/vectors 
$\Delta$, $\vec{\gamma}$ by $2\times 2$ matrices
(we carry this out for any spatial dimension; the case $d=2$ will be needed in section~3 below for the {\sc ecga})
\BEQ \label{2.27}
\Delta \mapsto \wht{\Delta} := \left( \matz{\Delta}{\Delta'}{0}{\Delta}\right) \;\; , \;\;
\vec{\gamma} \mapsto \wht{\vec{\gamma}} := 
\left( \matz{\vec{\gamma}}{\vec{\gamma'}}{\vec{\gamma''}}{\vec{\gamma}}\right)
\EEQ
where we already used that one of the two matrices can without restriction of the generality be assumed to have a 
(non-diagonalisable) Jordan form.\footnote{This discussion is quite analogous to the one 
which applies to the logarithmic
representations of the `ageing' sub-algebra of the Schr\"odinger algebra 
(without time-translations) \cite{Henkellog}; 
the two scaling dimensions $x,\xi$ used
therein and their matrix generalisations play the same r\^oles as 
$\Delta,\gamma$ in the CGA studied here.} In order to find
the most general admissible form, we write the above representations (\ref{gcageneratrs}) 
of the CGA as follows (and include all terms which describe the
transformation of the scaling operators), with $n\in\mathbb{Z}$
\BEA
{X^n} &=& -t^{n+1}\partial_t -(n+1)t^n \vec{x}\cdot\vec{\partial}_{\vec{x}} 
-(n+1)t^n \left(\matz{\Delta}{\Delta'}{0}{\Delta}\right) 
        -n(n+1)t^{n-1}\left(\matz{\vec{\gamma}}{\vec{\gamma'}}{\vec{\gamma''}}{\vec{\gamma}}\right)\cdot\vec{x} 
	\nonumber \\
{\vec{Y}^n} &=& -t^{n+1}\vec{\partial}_{\vec{x}} -(n+1) t^n 
\left(\matz{\vec{\gamma}}{\vec{\gamma'}}{\vec{\gamma''}}{\vec{\gamma}}\right) \label{2.28} \\
R &=& - \epsilon_{ij} x_i \partial_j - \epsilon_{k\ell} \gamma_k \frac{\partial}{\partial \gamma_{\ell}}
- \epsilon_{k\ell} \gamma'_k \frac{\partial}{\partial \gamma'_{\ell}}
- \epsilon_{k\ell} \gamma''_k \frac{\partial}{\partial \gamma''_{\ell}}  \nonumber
\EEA
(with $\partial_j := \partial/\partial x_j$) such that the non-vanishing commutators become
\BEA
{} [{X^n}, {X^m}] &=& 
(n-m) {X^{n+m}}+(n+1)(m+1)(m-n) t^{n+m-1} \Delta' \vec{\gamma''}\cdot\vec{x} \left(\matz{1}{0}{0}{-1}\right) 
\nonumber \\
{} [{X^n}, {\vec{Y}^m}] &=& 
(n-m) {\vec{Y}^{n+m}}+(n+1)(m+1) t^{n+m} \Delta' \vec{\gamma''} \left(\matz{1}{0}{0}{-1}\right) 
\EEA
and $[Y^n_{i},R]=-\epsilon_{i\ell} Y^n_{\ell}$. 
In order to recover the commutators (\ref{2.18}) of the CGA, we must have
\BEQ
\Delta' \,\vec{\gamma''} = \vec{0}
\EEQ
Hence, either $\Delta'=0$ such that the matrix 
$\wht{\Delta}=\Delta \left(\matz{1}{0}{0}{1}\right)$ and $\wht{\vec{\gamma}}$ 
is either diagonalisable (which would give a pair of non-logarithmic representations) 
or else it has a Jordan form where one can always
arrange for $\vec{\gamma''}=\vec{0}$. Alternatively, we have directly $\vec{\gamma''}=\vec{0}$. 
Therefore, {\em one can always set $\vec{\gamma''}=\vec{0}$ in (\ref{2.27}).} 

In summary without loss of generality, we can repeat eq.~(\ref{2.14}) by admitting as the most general case states :
\bea\label{eigenGCA2}\begin{split}
&X^0|\Delta,\gamma;1\rangle=0,
\;\;\;\;\;\;\;\;\;\;\;\;\;\;\;\;\;\;\;\;\;\;\;\;\;\;X^0|\Delta,\gamma;2\rangle=\Delta^\prime|\Delta,\gamma;1\rangle 
\cr& Y^0|\Delta,\gamma;1\rangle=0,
\;\;\;\;\;\;\;\;\;\;\;\;\;\;\;\;\;\;\;\;\;\;\;\;\;\;Y^0|\Delta,\gamma;2\rangle=\gamma^\prime|\Delta,\gamma;1\rangle
\end{split}\eea
In the language of nilpotent variables, we  define an eigenstate $|\widetilde{\Delta},\widetilde{\gamma}\rangle$ where
\bea\label{nilpotentdelta}
\widetilde{\Delta}\;=\;\Delta+\Delta^{\prime}\theta,\;\;\;\;\;\;\;\;\;\;\;\;\;\;
\widetilde{\gamma}\;=\;\gamma+\gamma^{\prime}\theta.
\eea
Equation (\ref{eigenGCA2}) then reads as:
\bea
X^0|\widetilde\Delta,\widetilde\gamma;2\rangle=\Delta^\prime\theta|\widetilde\Delta,\widetilde\gamma;1\rangle, \;\;\;\;\;\;\;\;\;\;\;\;Y^0|\widetilde\Delta,\widetilde\gamma;2\rangle=\gamma^\prime\theta|\widetilde\Delta,\widetilde\gamma;1\rangle.
\eea
Now we follow on to calculate two-point functions:
\bea\label{twopointlgca}
F(x_1,t_1,\theta_1;x_2,t_2,\theta_2 )=\langle\widetilde\Delta_1,\widetilde\gamma_1|\phi_1(x_1,t_1)\phi_2(x_2,t_2)|\widetilde\Delta_2,\widetilde\gamma_2\rangle 
\eea
Before going further let's redefine our parameters such that for the general variable $w$ we set:
\bea
w=w_1-w_2\;\;\;\;\;\;\;\;\;\;\;\;\;\;\;w^+=w_1+w_2.
\eea
For instance:
\bea\label{reparametrization}\begin{split}
&t=t_!-t_2 \;\;\;\;\;\;\;\;\;\;\;\;\;\;\;\;\;\;\;\;\;\;t^+=t_1+t_2
\cr&\theta=\theta_1-\theta_2\;\;\;\;\;\;\;\;\;\;\;\;\;\;\;\;\;\;\;\theta^+=\theta_1+\theta_2
\cr&\Delta=\Delta_1-\Delta_2\;\;\;\;\;\;\;\;\;\;\;\;\;\;\;\Delta^+=\Delta_1+\Delta_2
\cr&etc.
\end{split}\eea
Two-point functions should be invariant under the Ward identities arising out of CGA elements 
${{X^{-1},X^0,X^1,Y^{-1},Y^0,Y^1}}$. First, since 
$F$ must be invariant under space- and time-translation, it must be a function merely 
of $t$ and $x$ an not of $t^+$ and $x^+$.\footnote{Implicitly, $\Delta'$ and 
$\gamma'$ are assumed to have the same value for both scaling operators.} 
Invariance under $Y^0$ is expressed as
\bea
(t_1\partial_{x_1}+\gamma_1+\gamma^{\prime}_1\theta_1+t_2\partial_{x_2}+\gamma_2+\gamma^{\prime}_2\theta_2)F=0,
\eea
which reduces to:
\bea
(t\partial_x+\gamma^{+}+\gamma^\prime_1\theta_1+\gamma^{\prime}_2\theta_2)F=0
\eea
restricting $F$ to:
\bea
F=e^{-(\gamma^{+}+\gamma^{\prime}_1\theta_1+\gamma^\prime_2\theta_2)
\frac{x}{t}}[g_0(t)+g_1(t)\theta_1+g_2(t)\theta_2+g_3(t)\theta_1\theta_2].
\eea
Now, invariance under $Y^1$ gives:
\bea
t^+(t\partial_x+\gamma^++\gamma^\prime_1\theta^1+\gamma^\prime_2\theta^2)F
+t(\gamma+\gamma^\prime_1 \theta_1-\gamma^\prime_2\theta^2)F=0.
\eea
So, we find that
\bea
\gamma=0,\;\;\;\;\;\;\;\;\;\;\;\;\;g_0(t)=0,\;\;\;\;\;\;\;\;\;\;\;\;\gamma^\prime_1g_2 =\gamma^\prime_2g_1,
\eea
reducing $F$ to:
\bea\label{F3}
F=e^{-(\gamma^{+}+\gamma^\prime_1\theta_1+\gamma^\prime_2\theta_2)\frac{x}{t}}
\left[w(t)\gamma^\prime_1\theta_1+w(t)\gamma^\prime_2\theta_2+g_3(t)\theta_1\theta_2\right]\,
\delta_{\gamma_1,\gamma_2}.
\eea
in which $w(t)=g_1(t)/\gamma^\prime_1$. Now let's look at $X^0$ which appears as
\bea
(t\partial_t+x\partial_x+\Delta^++\Delta^\prime_1\theta_1+\Delta^\prime_2\theta_2)F=0,
\eea
Inserting $F$ from (\ref{F3}) in the above equation leads to:
\bea\begin{split}
&(t\partial_t+\Delta^+)g_1(t)=0,
\cr&(t\partial_t+\Delta^+)g_3(t)+\Delta^\prime _1\gamma^\prime_2w(t)+\Delta^\prime _2\gamma^\prime_1w(t)=0,
\end{split}\eea
which results in
\bea\begin{split}
&w(t)=at^{-\Delta_+},
\cr&g_3(t)=t^{-\Delta_+}(b-a(\Delta^\prime _1\gamma^\prime_2+\Delta^\prime _2\gamma^\prime_1)\ln{|t|}).
\end{split}\eea
Action of $X^1$ gives nothing new but super-selection rules:
\bea
\Delta=0,\;\;\;\;\;\;\;\;\;\;\;\;\Delta^\prime _1\gamma^\prime_2=\Delta^\prime _2\gamma^\prime_1.\eea
This constraint will appear in $G_{12}$ and $G_{21}$. Since under exchange $1\leftrightarrow 2$ we have $G_{12}\leftrightarrow G_{21}$ we can renormalize $\phi$ in a way to arrange for a perfect symmetry under exchange of the scaling operators, such that
$\langle \phi_1\psi_2\rangle = G_{12}\stackrel{!}{=}G_{21}=\langle \psi_1\phi_2\rangle$. This leads to the stronger constraints:
\bea
\Delta^\prime _1=\Delta^\prime _2, \;\;\;\;\;\;\;\;\;\;\;\;\gamma^\prime_1=\gamma^\prime_2.
\eea
So, the final result is 
\bea\label{F}
F=e^{-(2\gamma_1+\gamma^{\prime}\theta^+)\frac{x}{t}}\left[at^{-2\Delta_1}\theta^+ 
+t^{-2\Delta_1}(b-2a\Delta^\prime \ln{|t|})\theta^+\theta^+\right]
\,\delta_{\gamma_1,\gamma_2}\delta_{\Delta_1,\Delta_2}
\delta_{\Delta^\prime _1,\Delta^\prime _2}\delta_{\gamma^\prime_1,\gamma^\prime_2}.
\eea
Expanding both sides of (\ref{twopointlgca}) in terms of nilpotent variables, 
we find the two-point functions for logarithmic primaries of CGA
\bea \label{2.48}\begin{split}
&\langle\phi_{1}\phi_{2}\rangle\;=\;0,&\cr&
\langle\phi_{1}\psi_{2}\rangle\;=\;a e^{-2\gamma_{1}\frac{x}{t}}\, t^{-2\Delta_{1}}\delta_
{\Delta_{1},\Delta_{2}}\delta_{\Delta^\prime _1,\Delta^\prime _2}\delta_{\gamma_1,\gamma_2}\delta_{\gamma^\prime_1,\gamma^\prime_2},&\cr&
\langle\psi_{1}\psi_{2}\rangle\;=\;e^{-2\gamma_{1}\frac{x}{t}}\, t^{-2\Delta_{1}}
\left[-2a\Delta^\prime_1\ln|t|-2a\;\gamma^{\prime}\frac{x}{t}+b\right]
\delta_
{\Delta_{1},\Delta_{2}}\delta_{\Delta^\prime _1,\Delta^\prime _2}\delta_{\gamma_1,\gamma_2}\delta_{\gamma^\prime_1,\gamma^\prime_2}.
\end{split}\eea 
One needs to notice that since $\phi\phi$ term is equal to zero, then we can re-scale $\phi_1$ so that 
$\Delta^\prime_1=\Delta^\prime_2$ and thereby $\gamma^\prime_1=\gamma^\prime_2$. 
These results can be obtained as well by contraction from a LCFT$_2$ 
where both chiral components have logarithmic partners.

Since we wrote the generators in (\ref{2.28}) in a arbitrary number of space dimensions $d$, 
it is now straightforward to write
down the extension of (\ref{2.48}) to $d+1$ dimensions
\BEA
\langle \phi_1 \phi_2\rangle(t,\vec{x}) &=& 0 \nonumber \\
\langle \phi_1 \psi_2\rangle(t,\vec{x}) &=& a |t|^{-2\Delta_1} e^{-2\vec{\gamma}_1\cdot\vec{x}/t} \: 
\delta_{\Delta_1,\Delta_2}\delta_{\vec{\gamma}_1,\vec{\gamma}_2}\, 
\delta_{\Delta'_1,\Delta'_2}\delta_{\vec{\gamma'}_1,\vec{\gamma'}_2}
\label{eq:cga_2p} \\
\langle \psi_1 \psi_2\rangle(t,\vec{x}) &=& |t|^{-2\Delta_1} e^{-2\vec{\gamma}_1\cdot\vec{x}/t} \,
\left[ b -2a \frac{\vec{x}}{t}\cdot\vec{\gamma'}_1 -2a \Delta_1' \ln|t|  \right] \:
\delta_{\Delta_1,\Delta_2}\delta_{\vec{\gamma}_1,\vec{\gamma}_2}\, 
\delta_{\Delta'_1,\Delta'_2}\delta_{\vec{\gamma'}_1,\vec{\gamma'}_2}
\nonumber
\EEA
with a manifest invariance under the spatial rotations (\ref{2.28}).\footnote{In principle, the constants $a,b$ can depend on the
vectors $\vec{\gamma}_1$ and $\vec{\gamma}_1'$. Rotation-invariance then implies 
$a=a(\vec{\gamma}_1^2,{\vec{\gamma}_1'}^2,\vec{\gamma}_1\cdot\vec{\gamma}_1')$ and analogously for $b$.} 
We also list explicitly the several super-selection rules, 
as they apply to the non-vanishing elements of the matrices $\wht{\Delta}$ and $\wht{\vec{\gamma}}$. 


\section{Exotic CGA}

The {\em exotic} CGA ({\sc ecga}) is a centrally extended CGA in $2+1$ dimensions. The generators
$P,K,F$ now become 2-dimensional vectors $\vec{P},\vec{K},\vec{F}$ (or equivalently $Y^n$ is replaced by $\vec{Y}^n$) such that 
the immediate extension of the commutators (\ref{2.18}) is centrally extended by the 
nontrivial commutators \cite{Lukierski1,Lukierski2}:
\bea
[K_i,K_j]=\Xi\epsilon_{ij},\;\;\;\;\;\;\;\;\;\;\;\;\;\;\;\;[P_i,F_j]=-2\Xi\epsilon_{ij} \;\;\;;\;\;\; i,j=1,2,
\eea
$\epsilon_{ij}$ are the elements of the totally antisymmetry matrix {\small$\weps=\left(\matz{0}{1}{-1}{0}\right)$} 
and and $\epsilon_{12}=1$. 
For realising the central charge, following \cite{Tachikawa} one may invoke two operators $\chi_i$ such that 
\bea \label{3.2}
[\chi_i,\chi_j]=\Xi\epsilon_{ij}\;\;\;\;\;\;\;\;\;\;\;\;\;\;\;\;[\chi_i,\Xi]=0.
\eea
Since $\Xi$ is central, one may represent it by its eigenvalue $\Xi =\xi$. 
The {\sc ecga} generators read:
\bea\label{nonrapidity}\begin{split}
&H=-\partial_t,\;\;\;\;\;\;\;\;\;\;\;\;\;\;\;D=-x_i\partial_i-t\partial_t,\;\;\;\;\;\;\;\;\;\;\; 
C=-2tx_i\partial_i-t^2\partial_t-2x_i\chi_i, \cr&
P_i=-\partial_i,\;\;\;\;\;\;\;\;\;\;\;\;\;\;\;K_i=-t\partial_i-\chi_i,\;\;\;\;\;\;\;\;\;\;\; \;\;
F_i=-t^2\partial_i-2t\chi_i-2x_j\epsilon_{ij}\xi ,\cr&J=-\epsilon_{ij}x_i\partial_j-\frac{1}{2\xi}\chi_i\chi_i,
\end{split}\eea
Here, the generator $J$ of rotations was explicitly included as well. Its commutators with the other generators of the
{\sc ecga} read
\BEA
& & \left[ J, H \right] \:=\: \left[ J, D \right] \:=\: \left[ J, C \right] \:=\: 0 \nonumber \\
& & \left[ J, \vec{P} \right] \:=\: \weps \vec{P} \;\; , \;\;
\left[ J, \vec{K} \right] \:=\: \weps \vec{K} \;\; , \;\;\left[ J, \vec{F} \right] \:=\: \weps \vec{F}
\EEA

Martelli and Tachikawa \cite{Tachikawa} obtain the above algebra by making a contraction of the $2+1$ dimensional conformal algebra where spin has been taken into account. In other words they start by:(with $ \mu $ and $\nu ={0,1,2}$)
\bea\label{ccc}\begin{split}
&\widetilde M_{\mu\nu}=-x_\mu\partial\nu+x_\nu\partial_\mu-\Sigma_{\mu\nu},\;\;\;\;\;\;\;\;\;\;\;\;\;\;\;\;\;\;\;\;\;\;\;\;\widetilde P_\mu=-\partial_\mu, \cr&\widetilde K_\mu=-x^\nu x_\nu\partial_\mu+2x_\mu x^\nu\partial_\nu+2x^\nu\Sigma_{\mu\nu}\;\;\;\;\;\;\;\;\;\;\;\;\;\;\widetilde D=-x^\mu\partial_\mu,
\end{split}\eea
where $\Sigma_{\mu\nu}$ is the spin. Now under the contraction limit (\ref{contraction}) and redefining operators as:
\bea\begin{split}
&P_i=\frac{\widetilde P_i}{c},\;\;\;\;\;\;\;\;\;\;\;\;\;H=P_0\;\;\;\;\;\;\;\;\;\;\;\;\;K_i=\frac{\widetilde M_{0i}}{c} \cr&D=\widetilde D\;\;\;\;\;\;\;\;\;\;\;\;\;\;F_i=\frac{\widetilde K_i}{c}\;\;\;\;\;\;\;\;\;\;\;\;\; C=-\widetilde K_0,
\cr&\chi_i=\frac{\Sigma_{0i}}{c} \;\;\;\;\;\;\;\;\;\;\;\;\xi=\frac{\Sigma_{21}}{c^2};\;\;\;\;\;\;\;\;\;\;\;\;\;J=\widetilde M_{12}+\frac{1}{2\xi c^2}(\Sigma_{0i}\Sigma_{0i})+\Sigma_{12}
\end{split}\eea
they end up with (\ref{nonrapidity}). Clearly, the operators $\chi_i$ and the central charge $\xi$  are remnants of the spin components.

The operators $\chi_i$ and the central charge $\xi$ 
can be realised explicitly via an auxiliary space with coordinates $\nu_1,\nu_2$:
\BEQ \label{3.5}
\chi_i=\partial_{\nu_i}-\frac{1}{2}\epsilon_{ij}\nu_j\xi
\EEQ
Alternatively, one may use instead of $J$ a more natural-looking generator of infinitesimal rotations, including its action on the
auxiliary space
\BEQ
R := - \epsilon_{ij} x_i \partial_{x_j} -- \epsilon_{ij} \gamma_i \partial_{\gamma_j} - \epsilon_{ij} \nu_i\partial_{\nu_j}
\EEQ
which obeys the same algebraic properties as the generator $J$. \\

In the above realisation, one expects the operators $D$ and $\Xi$ to have simultaneous eigenvectors since they commute, which is the primary state we use to construct the correlators. In order to include the rapidities as well, and to simplify the 
computation of two-point functions, we include all those
terms which describe the transformation of the scaling operators into the generators. 
Then the action of all generators
on two-point functions simply vanishes. The important Bargman super-selection rule of the central charge
$\xi^{[2]}=\xi_1 + \xi_2=0$ follows. This is completely analogous to the treatment of the central charge in the
Schr\"odinger algebra in section~2.2.
In this new representation, the generators of the {\sc ecga} read 
(those given in \cite{Tachikawa,Cherniha10} are included as special cases)
\bea\label{rapidity}\begin{split}
&H=X^{-1}=-\partial_t,\;\;\;\;\;\;\;\;\;\;\;\;\;\;\;\;\;\;\;\;\;\;\;\;\;\;\;\;\;\;\;\;\;
D=X^0=-x_i\partial_i-t\partial_t-\Delta,\;\;\;\;\;\;\;\;\;\;\; \cr&
C=X^1=-2tx_i\partial_i-t^2\partial_t-2\Delta t-2x_i\partial_{\nu_i}+\epsilon_{ij}\nu_jx_i\xi -2\gamma_i x_i, \cr&
P_i=Y^{-1}_i=-\partial_i,\;\;\;\;\;\;\;\;\;\;\;\;\;\;\;\;\;\;\;\;\;\;\;\;\;\;\;\;\;\;\;\;\;
K_i=Y^0_i=-t\partial_i-\partial_{\nu_i}+\frac{1}{2}\epsilon_{ij}\nu_j\xi -\gamma_i,\;\;\;\;\;\;\;\;\;\;\;\;\;\; \;\;\cr&
F_i=Y^1_i=-t^2\partial_i-2t\partial_{\nu_i}+t\epsilon_{ij}\nu_j\xi-2x_j\epsilon_{ij}\xi-2t\gamma_i
\end{split}\eea
This set is to be completed by a rotation generator for which we shall choose either $J$ or $R$. 
Two-point functions may now be derived explicitly from (\ref{rapidity}). We observe that rotation invariance under the action of 
the generator $R$ leads to a different result than requiring co-variance under the rotation
generator $J$ from (\ref{nonrapidity}), as used in \cite{Tachikawa}. In what follows,
we distinguish these two cases and speak of `{\em $J$-invariance}' if $J$ is used along with the generators of (\ref{rapidity}) and 
of `{\em $R$-invariance}' when $R$ is used.

Quite analogously to Schr\"odinger-invariance treated above, the generators (\ref{rapidity}) 
are dynamical symmetries of the wave equation ${\cal S}\phi=0$, where
the Schr\"odinger operator is
\BEQ
{\cal S} := -\xi \partial_t + \epsilon_{ij} (\chi_i +\gamma_i) \partial_j 
=- \xi \partial_t + \left( \vec{\chi} +\vec{\gamma}\right)  \weps \vec{\partial}_{\vec{x}}
\EEQ
The only generators of the {\sc ecga} (\ref{rapidity}) which do not commute with $\cal S$ are $D$ and $C$: 
\BEQ
\left[ {\cal S}, D \right] = -{\cal S} \;\; , \;\; \left[ {\cal S}, C \right] = -2t{\cal S} -2\xi (\Delta -1) 
\EEQ
Rotation-invariance holds as well: $[{\cal S},J]=[{\cal S},R]=0$. 
Hence one has a dynamical symmetry on the space of solutions of the equation ${\cal S}\phi=0$ where $\Delta=\Delta_{\phi}=1$,
consistent with the known unitary bound $\Delta\geq 1$ \cite{Tachikawa}. This illustrates again the importance of these non-trivial
central extensions for the existence of non-trivial non-conformal wave equations.

\subsection{Non-logarithmic two-point functions}
As a first step towards the logarithmic two-point functions from the {\sc ecga}, 
we begin with the non-logarithmic case. This
was done first by Martelli and Tachikawa \cite{Tachikawa}, 
but only for vanishing rapidities $\vec{\gamma}_i=\vec{0}$. 
It is one of the aims of this
section to allow for $\vec{\gamma}_i \ne \vec{0}$ 
and to analyse systematically the possible constraints. The two-point function is defined as 
\BEQ \label{rapiditytwo}
F:=F(t_1,t_2;\vec{x}_1,\vec{x}_2;\vec{\nu}_1,\vec{\nu}_2)
=\left\langle\phi_1(t_1,\vec{x}_1,\vec{\nu}_1)\phi_2(t_2,\vec{x}_2,\vec{\nu}_2)\right\rangle.
\EEQ
We shall use throughout the variables as defined in (\ref{reparametrization}) 
and apply the {\sc ecga}-Ward identities derived from
(\ref{rapidity}) to $F$. 
Space- and time-translation-invariance restrict $F$ to be a function of $t=t_1-t_2$ and $\vec{x}=\vec{x}_1-\vec{x}_2$.  
The invariance under the central charge
$\Xi$ gives the important Bargman super-selection rule
\BEQ \label{Barg_nl}
\xi_1 + \xi_2 = 0
\EEQ
Invariance under the dilatations $D$ and the
two generalised Galilei-transformations $\vec{K}$ gives 
\BEA
\left( -t\partial_t -\vec{x}\cdot\vec{\partial}-\Delta_1-\Delta_2\right) F &=& 0 \label{X0} \\
\left( -t\vec{\partial} -\vec{\gamma}_1 -\vec{\gamma}_2 - \vec{\chi}_1 - \vec{\chi}_2 \right) F &=& 0 \label{Y0}
\EEA
Rather than using these to parametrise immediately an explicit scaling form, 
we prefer to use these identities first to simplify the
remaining conditions and especially to derive the constraints the two-point function $F$ must obey. 
Therefore, we next rewrite the
Ward identity of the two generators $\vec{F}$, which gives.
$\left(-t^2\vec{\partial}-2t\left(\vec{\chi}_1+\vec{\gamma}_1\right)-2\xi_1 \weps \vec{x}\right)F=0$ 
and where the eqs.~(\ref{Barg_nl},\ref{Y0}) have been used. 
Using again (\ref{Y0}), this is further simplified to
\BEQ \label{Y1}
\left( -t \left( \vec{\chi}_1 - \vec{\chi}_2 + \vec{\gamma}_1 - \vec{\gamma}_2\right) -2\xi_1 \weps\vec{x}\right) F=0
\EEQ
Similarly, invariance under $C$ gives $\left(-t^2\partial_t -2t\vec{x}\cdot\vec{\partial}-2\Delta_1 t
-2(\vec{\chi}_1+\vec{\gamma}_1)\cdot\vec{x}\right)F=0$, 
where eqs.~(\ref{Barg_nl},\ref{X0},\ref{Y0}) have been used. Applying (\ref{X0}),
again, leads to
\BEQ \label{X1}
\left( -t\vec{x}\cdot\vec{\partial}-\left(\Delta_1 -\Delta_2\right)t 
-2\left( \vec{\chi}_1 +\vec{\gamma}_1\right)\cdot\vec{x} \right) F=0
\EEQ
Eqs.~(\ref{X0},\ref{Y0},\ref{Y1},\ref{X1}) 
contain the complete available information on the shape of the two-point function $F$, up to rotation-invariance, to be discussed below.  

Multiplying (\ref{Y1}) with $\vec{x}$, one has 
\BEQ \label{3.13}
\vec{x}\cdot \left( \vec{\chi}_1 +\vec{\gamma}_1 \right) F 
= \vec{x}\cdot \left( \vec{\chi}_2 +\vec{\gamma}_2 \right) F
\EEQ
such that comparison of eqs.~(\ref{Y0},\ref{X1}) leads to $(\Delta_1-\Delta_2)t F=0$. 
Hence,  the constraint $\Delta_1=\Delta_2$ 
follows.\footnote{For the non-exotic CGA, (\ref{Y1}) or (\ref{3.13}) would further imply $\vec{\gamma}_1=\vec{\gamma}_2$.} 

The remaining three independent equations can be further simplified via  the ansatz
\BEQ  \label{3.16}
F = t^{-2\Delta_1} e^{-(\vec{\gamma}_1+\vec{\gamma}_2)\cdot \vec{u}} f\left(\vec{u}, \vec{\nu}_1, \vec{\nu}_2\right) 
\;\; , \;\; \vec{u} := \vec{x}/t
\EEQ
which leads to the following two conditions
\BEQ
\left( \partial_{\vec{u}} +\vec{\chi}_1 + \vec{\chi}_2 \right) f = 0 \;\; , \;\;
\left( \vec{\chi}_1 - \vec{\chi}_2 + \vec{\gamma}_1 - \vec{\gamma}_2 + 2\xi_1 \weps \vec{u} \right) f =0
\EEQ
Only now, we use the explicit form (\ref{3.5}). Further, we introduce the new variables 
$\vec{\nu}^{\pm} := \demi \left( \vec{\nu}_1 \pm \vec{\nu}_2 \right)$ and 
write $f=f(\vec{u},\vec{\nu}^+,\vec{\nu}^-)$ such that
\BEQ
\left( \partial_{\vec{u}} + \partial_{\vec{\nu}^+} -\xi_1 \weps \vec{\nu}^- \right) f = 0 \;\; , \;\;
\left( \partial_{\vec{\nu}^-} -\xi_1 \weps \vec{\nu}^+ 
+2\xi_1 \weps \vec{u} +\vec{\gamma}_1 - \vec{\gamma}_2 \right) f =0 
\EEQ
The first of those is solved by the ansatz 
$f=\exp\left[\xi_1 \vec{u}\weps\vec{\nu}^-\,\right] \phi(\vec{w},\vec{\nu}^-)$, 
with $\vec{w} :=\vec{u}-\vec{\nu}^+$. The last function $\phi$ can be found from the equation
$\left(\partial_{\vec{\nu}^-}+\xi_1 \weps\vec{w} +\vec{\gamma}_1 -\vec{\gamma}_2\right)\phi=0$. Hence
\BEQ \label{3.19}
\phi\left(\vec{w},\vec{\nu}^-\right) = \phi_0\left(\vec{w}\right) 
\exp\left[ -\xi_1 \vec{\nu}^- \cdot \weps \vec{w} -\vec{\nu}^- \cdot(\vec{\gamma}_1 - \vec{\gamma}_2) \right]
\EEQ
where $\phi_0$ is an arbitrary differentiable function, which besides on $\vec{w}$, can in principle also depend on the parameter
vectors $\vec{\gamma}_{1,2}$. 

Summarising our results, we can write the explicit two-point function, with $\vec{u}=\vec{x}/t$
\BEQ\label{fifi}
\langle \phi_1 \phi_2\rangle = f_0\left(\vec{u}-\frac{\vec{\nu}_1+\vec{\nu}_2}{2}\right) |t|^{-2\Delta_1}\, 
e^{-(\vec{\gamma}_1+\vec{\gamma}_2)\cdot\vec{u} 
-\demi (\vec{\gamma}_1-\vec{\gamma}_2)\cdot(\vec{\nu}_1-\vec{\nu}_2)}\, 
e^{\xi_1 \vec{u}\wedge(\vec{\nu}_1-\vec{\nu}_2) +\demi\xi_1 \vec{\nu}_1\wedge\vec{\nu}_2}\,
\delta_{\Delta_1,\Delta_2} \delta_{\xi_1+\xi_2,0}
\EEQ
where the symmetry under exchange of position $1\leftrightarrow 2$ is taken into 
account.\footnote{The correlator obtained from the free-field solution
of ${\cal S}\phi=0$ is of this form, with $f_0=\mbox{\rm cste.}$ \cite{Tachikawa}.}  One also uses the notation of the skew product
$\vec{a}\wedge\vec{b} := \vec{a}\weps\vec{b}=\epsilon_{ij} a_i b_j$. 
If the rotation-invariance is taken into account  as well, the undetermined function written above as $f_0=f_0(\vec{w})$ becomes
\BEA
\mbox{\rm $J$-invariance} &:& f_0=f_0\left(\vec{\gamma}_1^2, \vec{\gamma}_2^2, \vec{\gamma}_1\cdot\vec{\gamma}_2, 
\vec{w}+\weps \left(\vec{\gamma}_1-\vec{\gamma}_2\right) (2\xi_1)^{-1} \right) 
\nonumber \\
\mbox{\rm $R$-invariance} &:& f_0=f_0\left(\vec{w}^2,\vec{\gamma}_1^2, \vec{\gamma}_2^2, 
\vec{w}\cdot\vec{\gamma}_1, \vec{w}\cdot\vec{\gamma}_2 \right)
\label{3.24}
\EEA
as is shown in appendix~B.

Remarkably, the {\sc ecga}-covariant two-point function no longer needs to obey the constraint 
$\vec{\gamma}_1 = \vec{\gamma}_2$ of the
rapidities which we had obtained in (\ref{eq:cga_2p}) for the non-exotic case. 


\subsection{Logarithmic Two-Point Functions}

We are finally prepared for the computation of the co-variant 
two-point functions in the logarithmic representation of the {\sc ecga},
which in the most simple case can be obtained formally from the representation 
(\ref{rapidity}) by making the replacements
\BEQ 
\Delta \mapsto \wht{\Delta} := \left( \matz{\Delta}{\Delta'}{0}{\Delta}\right) \;\; , \;\;
\vec{\gamma} \mapsto \wht{\vec{\gamma}} := \left( \matz{\vec{\gamma}}{\vec{\gamma'}}{\vec{0}}{\vec{\gamma}}\right)
\EEQ
see section~2.3. We seek the two-point functions
\BEQ \label{ecga_FGH}
F = \langle \phi_1 \phi_2\rangle \;\; , \;\;
G_{12} = \langle \phi_1 \psi_2\rangle \;\; , \;\;
G_{21} = \langle \psi_1 \phi_2\rangle \;\; , \;\;
H = \langle \psi_1 \psi_2\rangle
\EEQ
where the arguments are implicit. Surprisingly, it turns out that the non-modified contribution 
$F=\langle \phi_1\phi_2\rangle$ 
does not necessarily vanish, in contrast with the non-exotic CGA (\ref{eq:cga_2p}). 
Throughout, temporal and spatial translations-invariance and invariance under the central
charge $\Xi$ shall be used, such that all two-point functions depend on 
$t$ and $\vec{x}$ and the Bargman super-selection rule
(\ref{Barg_nl}) is valid. 

We now state the explicit result and refer to appendix~A for the details of the calculation. 
Two distinct cases must be recognised, namely 
\begin{enumerate}
\item \underline{\bf Case 1}: $\Delta_1'\ne0$ or $\Delta_2'\ne 0$ and $F=0$. 
This is the most direct extension of the logarithmic representations
of the non-exotic CGA. 
\item \underline{\bf Case 2}: $\Delta_1'=\Delta_2'= 0$ and $F\ne 0$. 
Here, only the rapidity matrices $\wht{\vec{\gamma}}_i$ will take a Jordan form, while $\wht{\Delta}=\Delta\,\wht{\bf 1}$ is diagonal. 
\end{enumerate} 
In what follows, we shall use the notations from section~3.1. \\

\noindent In \underline{\it Case 1}, we have $F=0$ and $G_{12}=G(t,\vec{x})=G(-t,-\vec{x})=G_{21}=:G$ such that
\BEA
G &=& |t|^{-2\Delta_1} 
e^{-(\vec{\gamma}_1+\vec{\gamma}_2)\cdot\vec{u} 
-\demi (\vec{\gamma}_1-\vec{\gamma}_2)\cdot(\vec{\nu}_1-\vec{\nu}_2)}\, 
e^{\xi_1 \vec{u}\wedge(\vec{\nu}_1-\vec{\nu}_2) +\demi\xi_1 \vec{\nu}_1\wedge\vec{\nu}_2}\,
g_0(\vec{w}) 
\nonumber \\
H &=& |t|^{-2\Delta_1} 
e^{-(\vec{\gamma}_1+\vec{\gamma}_2)\cdot\vec{u} 
-\demi (\vec{\gamma}_1-\vec{\gamma}_2)\cdot(\vec{\nu}_1-\vec{\nu}_2)}\, 
e^{\xi_1 \vec{u}\wedge(\vec{\nu}_1-\vec{\nu}_2) +\demi\xi_1 \vec{\nu}_1\wedge\vec{\nu}_2}\,
h(\vec{u},\vec{\nu}_1,\vec{\nu}_2) 
\label{ecga_case1} \\
h &=& h_0(\vec{w}) - g_0(\vec{w}) \left( 2\Delta_1' \ln|t| + \vec{u}\cdot\left(\vec{\gamma}_1'+\vec{\gamma}_2'\right)
+\demi\left(\vec{\nu}_1-\vec{\nu}_2\right)\cdot\left(\vec{\gamma}_1'-\vec{\gamma}_2'\right) \right)
\nonumber
\EEA
together with the abbreviations $\vec{u}=\vec{x}/t$ and 
$\vec{w} := \vec{u}-\demi\left(\vec{\nu}_1+\vec{\nu}_2\right)$ 
and the constraints $\Delta_1=\Delta_2$, $\Delta_1'=\Delta_2'$ and  $\xi_1+\xi_2=0$. The undetermined functions $g_0(\vec{w})$ 
and $h_0(\vec{w})$ still are subject to rotation-invariance, see below. \\

\noindent In \underline{\it Case 2}, we find the constraints $\Delta_1=\Delta_2$ and  $\xi_1+\xi_2=0$ and
\BEA
F &=& |t|^{-2\Delta_1} 
e^{-(\vec{\gamma}_1+\vec{\gamma}_2)\cdot\vec{u} 
-\demi (\vec{\gamma}_1-\vec{\gamma}_2)\cdot(\vec{\nu}_1-\vec{\nu}_2)}\, 
e^{\xi_1 \vec{u}\wedge(\vec{\nu}_1-\vec{\nu}_2) +\demi\xi_1 \vec{\nu}_1\wedge\vec{\nu}_2}\,
f_0(\vec{w}) 
\nonumber \\
G_{12} &=& |t|^{-2\Delta_1} 
e^{-(\vec{\gamma}_1+\vec{\gamma}_2)\cdot\vec{u} 
-\demi (\vec{\gamma}_1-\vec{\gamma}_2)\cdot(\vec{\nu}_1-\vec{\nu}_2)}\, 
e^{\xi_1 \vec{u}\wedge(\vec{\nu}_1-\vec{\nu}_2) +\demi\xi_1 \vec{\nu}_1\wedge\vec{\nu}_2}\,
g_{12}(\vec{u},\vec{\nu}_1,\vec{\nu}_2) 
\nonumber \\
G_{21} &=& |t|^{-2\Delta_1} 
e^{-(\vec{\gamma}_1+\vec{\gamma}_2)\cdot\vec{u} 
-\demi (\vec{\gamma}_1-\vec{\gamma}_2)\cdot(\vec{\nu}_1-\vec{\nu}_2)}\, 
e^{\xi_1 \vec{u}\wedge(\vec{\nu}_1-\vec{\nu}_2) +\demi\xi_1 \vec{\nu}_1\wedge\vec{\nu}_2}\,
g_{21}(\vec{u},\vec{\nu}_1,\vec{\nu}_2) 
\label{ecga_case2}\\
H &=& |t|^{-2\Delta_1} 
e^{-(\vec{\gamma}_1+\vec{\gamma}_2)\cdot\vec{u} 
-\demi (\vec{\gamma}_1-\vec{\gamma}_2)\cdot(\vec{\nu}_1-\vec{\nu}_2)}\, 
e^{\xi_1 \vec{u}\wedge(\vec{\nu}_1-\vec{\nu}_2) +\demi\xi_1 \vec{\nu}_1\wedge\vec{\nu}_2}\,
h(\vec{u},\vec{\nu_1},\vec{\nu}_2) 
\nonumber 
\EEA
where 
\BEA
g_{12} &=&  g_{0}(\vec{w}) - f_0(\vec{w}) 
\left( \vec{u}-\demi\left(\vec{\nu}_1-\vec{\nu}_2\right)\right)\cdot\vec{\gamma}_2' \nonumber \\
g_{21} &=&  g_{0}(\vec{w}) - f_0(\vec{w}) 
\left( \vec{u}+\demi\left(\vec{\nu}_1-\vec{\nu}_2\right)\right)\cdot\vec{\gamma}_1'  
\label{ecga_gh}\\
h      &=& h_0(\vec{w}) - g_0(\vec{w}) \left( \vec{u}\cdot\left(\vec{\gamma}_1'+\vec{\gamma}_2'\right)
+\demi\left(\vec{\nu}_1-\vec{\nu}_2\right)\cdot\left(\vec{\gamma}_1'-\vec{\gamma}_2'\right) \right)
\nonumber \\
& & +\demi f_0(\vec{w}) \left( \vec{u}+\demi\left(\vec{\nu}_1-\vec{\nu}_2\right)\right)\cdot\vec{\gamma}_1' \: 
\left( \vec{u}-\demi\left(\vec{\nu}_1-\vec{\nu}_2\right)\right)\cdot\vec{\gamma}_2'
\nonumber
\EEA

Finally, in both cases, rotation-invariance must be taken into account, see appendix~B for the details. If we use $R$-invariance, 
in both cases the functions $f_0(\vec{w})$, $g_{0}(\vec{w})$ 
and $h_0(\vec{w})$ are short-hand notations for undetermined functions of 9 
rotation-invariant combinations of $\vec{w}$, $\vec{\gamma}_{1,2}$ and
$\vec{\gamma}_{1,2}'$, for example
\BEQ \label{3.30}
f_0=f_0\left( \vec{w}^2,\vec{\gamma}_1^2,\vec{\gamma}_2^2,{\vec{\gamma}_1'}^2, 
{\vec{\gamma}_2'}^2,\vec{w}\cdot\vec{\gamma}_1,\vec{w}\cdot\vec{\gamma}_2,
\vec{w}\cdot\vec{\gamma}_1',\vec{w}\cdot\vec{\gamma}_2'\right)
\EEQ
and analogously for $g_0$ and $h_0$. 
We point out that in both cases, there is no constraint neither on the
$\vec{\gamma}_i$, nor the $\vec{\gamma}_i'$. Alternatively, if we use $J$-invariance we find that the $\vec{\gamma}$-matrices are diagonal, 
viz. $\vec{\gamma}_1'=\vec{\gamma}_2'=\vec{0}$. Then {\em only case 1 retains a logarithmic structure} and we have 
\BEA 
g_0=g_0\left( \vec{\gamma}_1^2,\vec{\gamma}_2^2,\vec{\gamma}_1\cdot\vec{\gamma}_2, 
\vec{w}+\weps \left(\vec{\gamma}_1-\vec{\gamma}_2\right) (2\xi_1)^{-1} \right)
\nonumber \\
h_0=h_0\left( \vec{\gamma}_1^2,\vec{\gamma}_2^2,\vec{\gamma}_1\cdot\vec{\gamma}_2, 
\vec{w}+\weps \left(\vec{\gamma}_1-\vec{\gamma}_2\right) (2\xi_1)^{-1} \right)
\label{3.31}
\EEA
So, the task is done and two-point functions of logarithmic representations of the {\sc ecga} have been worked out.


\section{Conclusions}

The exotic Galilean algebra corresponds to $d=2$ and $l=1$ case of $l$-Galilei algebras. 
This algebra arises as the singular limit of the conformal algebra when the speed of light tends to infinity. 
In other words it should describe the low velocity limit of conformal systems. However the low energy limit is often described by the Schr\"odinger algebra which is the $l=\frac{1}{2}$ case of $l$-Galilei algebras.  
This is rather paradoxical and the physical candidates for the realisation of CGA have proved hard to find. 

In this work, we analysed the generic form of co-variant two-point functions, for scalar and logarithmic representations
of conformal galilean algebra. The transformation of quasi-primary scaling operators under these algebras can be 
characterised in terms of a scaling dimension $\Delta$, a rapidity vector $\vec{\gamma}$ and in the case of the exotic
{\sc ecga} also in terms of a `mass' $\xi$. If one considers logarithmic representations, the scaling dimension and the
rapidities can acquire a matrix form. Restricting to the most simple case of two-component logarithmic representations, these
matrices have been shown to be {\em simultaneously} of a Jordan form
\BEQ \label{4.1}
\Delta \mapsto \wht{\Delta} = \left(\matz{\Delta}{\Delta'}{0}{\Delta}\right) \;\; , \;\;
\vec{\gamma} \mapsto \wht{\vec{\gamma}} = \left(\matz{\vec{\gamma}}{\vec{\gamma}'}{\vec{0}}{\vec{\gamma}}\right)
\EEQ
Qualitatively very different results were obtained for the non-exotic CGA and the exotic {\sc ecga}, as summarised in table~\ref{tab1}.
\begin{table}
\begin{center}\begin{tabular}{|l|c|clclc|l|} \hline
algebra & ~eq.~ & \multicolumn{5}{c|}{constraints} & \\ \hline 
CGA   & (\ref{eq:F_cga})  & $\Delta_1=\Delta_2$ & & $\vec{\gamma}_1=\vec{\gamma}_2$ & & & \\[0.2truecm]
L-CGA & (\ref{eq:cga_2p}) & $\Delta_1=\Delta_2$ & $\Delta_1'=\Delta_2'$             & $\vec{\gamma}_1=\vec{\gamma}_2$ & 
$\vec{\gamma}_1'=\vec{\gamma}_2'$ &  & \\[0.2truecm]
{\sc ecga}   & (\ref{fifi})       & $\Delta_1=\Delta_2$ &                         & & &  $\xi_1+\xi_2=0$  & \\[0.25truecm]
             & (\ref{ecga_case1}) & $\Delta_1=\Delta_2$ & $\Delta_1'=\Delta_2'$   & & & $\xi_1+\xi_2=0$ & R1 \\
L-{\sc ecga} & (\ref{ecga_case2}) & $\Delta_1=\Delta_2$ & $\Delta_1'=\Delta_2'=0$ & & & $\xi_1+\xi_2=0$ & R2 \\
	         & (\ref{ecga_case1}) & $\Delta_1=\Delta_2$ & $\Delta_1'=\Delta_2'$   & & $\vec{\gamma}_1'=\vec{\gamma}_2'=\vec{0}$ & $\xi_1+\xi_2=0$ & J  \\
\hline
\end{tabular}\end{center}
\caption[tab1]{Summary on the constraints obeyed by co-variant two-point functions in several variants of conformal galilean algebras. The 
first column indicates the non-exotic algebra CGA or the exotic {\sc ecga}, where a prefix `$\mbox{\rm L-}$' indicates a 
logarithmic representation. The equation labels refer to the explicit form of the two-point function, as derived in the text.
The various constraints on either scaling dimensions $\Delta$, rapidities $\vec{\gamma}$ or 
the Bargman super-selection rule on the `masses' $\xi$ are listed. 
The last three lines refer to the logarithmic representations of the exotic {\sc ecga}. 
Therein, the extra labels refer to the two distinct choices of the rotation generator: 
either $R$-invariance with the two distinct case 1 (R1) and case 2 (R2), 
or else $J$-invariance (J). \label{tab1}}
\end{table}

\begin{enumerate}
\item When considering the CGA, the extension to logarithmic representation essentially produced the expected generalisations of the
constraints on both the scaling dimension $\Delta$ and the rapidity $\vec{\gamma}$ also to the non-diagonal elements of
the corresponding matrices, according to (\ref{4.1}). In addition, the various two-point functions simply take up the same
kind of logarithmic prefactors, see eq.~(\ref{eq:cga_2p}), as one would have expected from logarithmic conformal or even
logarithmic Schr\"odinger-invariance, see eqs.~(\ref{eq:lconf},\ref{eq:lsch}). 

\item Therefore, by comparing the results (\ref{eq:lconf}) of logarithmic conformal invariance and 
(\ref{eq:lsch}) of logarithmic Schr\"odinger-invariance, 
one might have believed that going over to the {\sc ecga} would merely lead to the naturally expected Bargman
super-selection rule $\xi_1+\xi_2=0$, which would be the analogue of non-relativistic mass conservation in Schr\"odinger-invariant
systems. Remarkably, it turned out that the form of the co-variant two-point functions in the
exotic {\sc ecga} is considerably richer. 

\item For scalar representations, the presence of extra internal dimensions needed to
realise the extra exotic structure releases the constraint on the rapidities $\vec{\gamma}_i$ of the two scaling operators. 
A finer difference arise through the possibility of two distinct choices for the generator of rotation, labelled
$J$ and $R$, and referred to as `{\em $J$-invariance}' and `{\em $R$-invariance}'. The precise shape of a last undetermined
scaling function $f_0$ depends on whether $J$-invariance or $R$-invariance is assumed, see eq.~(\ref{3.24}). 

\item Stronger qualitative differences appear in the logarithmic representations of the {\sc ecga}. For $R$-invariance, 
two distinct cases emerge. The first one, labelled R1 in table~\ref{tab1}, 
keeps the conventional result that the two-point function $F=\langle \phi \phi\rangle=0$
of the partner vanishes. 
But if the matrices $\wht{\Delta}$ are diagonal, a new case arises, labelled R2 in table~\ref{tab1}, where $F\ne 0$ and new additional
terms in the remaining two-point functions $\langle \phi\psi\rangle$ and $\langle \psi\psi\rangle$ arise. In both cases, the
remaining scaling functions are of the generic form (\ref{3.30}).   
On the other hand, for $J$-invariance, labelled J in table~\ref{tab1}, 
the rapidity matrices $\wht{\vec{\gamma}}$ are forced to be diagonal, such that
the logarithmic terms reduce to those known from the non-exotic CGA. 
Here, only case 1 retains a logarithmic structure and the form of the remaining scaling functions is given in (\ref{3.31}). 

\item 

What has been referred to in the literature as ``logarithmic" conformal field theory, 
uses in fact reducible but non-decomposable representations of the conformal algebra. 
In all cases known so far, the correlators also acquired a logarithmic term as well as power-law-dependence on distance, which
motivated the name `{\em logarithmic}'. Here, a first example has been found (case R2 of the L-{\sc ecga} in table~\ref{tab1}) 
where a reducible but non-decomposable representation 
does not lead to explicit logarithms in the two-point functions.

\end{enumerate} 

The present study looked at co-variant two-point functions of conformal galilean algebras from an abstract point of view. 
We hope that the results presented here will turn out to be helpful in identifying specific physical model with one of them
as a dynamical symmetry. We hope to come back to this in future work.

\section*{Acknowledgments}
We are grateful to S. Moghimi-Araghi and A. Naseh for discussions. 
Two of us (MH and SR) would like to thank the organisers of the ADCFTA at the IHP Paris 2011, 
for warm hospitality, where this work was initiated. MH thanks the organisers of the Symposium 
`Models from statistical mechanics in applied sciences' at Warwick University 2013 for warm hospitality and acknowledges 
partial support from the Coll\`ege Doctoral franco-allemand Nancy-Leipzig-Coventry
(Syst\`emes complexes \`a l'\'equilibre et hors \'equilibre) of UFA-DFH.  
SR would like to thank warm hospitality at Van der Waals-Zeeman Instituut, University of Amsterdam,Nederlands and Groupe de Physique  Statistique, Institut Jean Lamour, University of Lorraine, France where part of this work were carried out.


\appsection{A}{Calculational details}

The results (\ref{ecga_case1}) and (\ref{ecga_case2},\ref{ecga_gh}) of the main text, 
respectively, are derived. 

Starting from the definitions (\ref{ecga_FGH}), the function $F$ 
was already found in section~3.1. As we shall see that $F\ne 0$ may occur,
we revert to the standard formulation of LCFT. 
Temporal and spatial translation-invariance and the Bargman super-selection rule
$\xi_1+\xi_2=0$ are obvious. 

We begin with the two-point function $G_{12}=\langle \phi_1 \psi_2\rangle$. 
Co-variance under the generators $X_0, Y_0, Y_1$ and $X_1$, 
via calculations totally analogous to the ones presented in section~3.1, lead to the conditions
\BEA
\left( -t \partial_t-\vec{x}\cdot\vec{\partial}-\Delta_1-\Delta_2\right) G_{12} - \Delta_2' F &=& 0 \nonumber \\
\left( -t\vec{\partial} - \vec{\gamma}_1 -\vec{\gamma}_2 
- \vec{\chi}_1 -\vec{\chi}_2 \right) G_{12} - \vec{\gamma}_2'F &=& 0 
\nonumber \\
\left( -t\left(\vec{\gamma}_1-\vec{\gamma}_2 +\vec{\chi}_1 -\vec{\chi}_2\right) 
-2\xi_1 \weps \vec{r} \right) G_{12} +t\vec{\gamma}_2'F &=& 0
\label{A1} \\
\left( -t\vec{r}\cdot\vec{\partial} -t(\Delta_1-\Delta_2) 
-2\vec{r}\cdot\left(\vec{\chi}_1+\vec{\gamma}_1\right)\right)G_{12}
+t\Delta_2' F &=& 0 
\nonumber 
\EEA
If $F\ne 0$, then from section~3.1 we have $\Delta_1=\Delta_2$. 
Otherwise, if $F=0$, the above conditions are then identical to those
treated in section~3.1, so that again $\Delta_1=\Delta_2$ follows. 
Hence, {\it we always have the constraint $\Delta_1=\Delta_2$.} Next, 
multiply the 3$^{\rm rd}$ eq.~(\ref{A1}) with $\vec{x}$. 
On the other hand, simplify the 4$^{\rm th}$ eq.~(\ref{A1}) by using again
$Y_0$-covariance. This gives the two simultaneous conditions
\BEA
t \vec{x} \cdot \left( \vec{\chi}_1-\vec{\chi}_2 +\vec{\gamma}_1 -\vec{\gamma}_2 \right) G_{12} 
-t \vec{x}\cdot\vec{\gamma}_2' F &=& 0
\\
\vec{x} \cdot \left( \vec{\chi}_1-\vec{\chi}_2 +\vec{\gamma}_1 -\vec{\gamma}_2 \right) G_{12} 
-\vec{x}\cdot\vec{\gamma}_2' F -t\Delta_2' F &=& 0
\nonumber
\EEA
which requires that
\BEQ
t^2 \Delta_2' F = 0
\EEQ
Similarly, if we consider the other mixed two-point function $G_{21}=\langle\psi_1\phi_2\rangle$, 
we find $t^2\Delta_1' F=0$. Therefore, {\it the following
cases must be distinguished:\footnote{We leave out here distributional contributions $F\sim \delta(t), \delta'(t)$.}
\begin{enumerate}
\item \underline{\bf Case 1}: Let $\Delta_1'\ne 0$ or $\Delta_2'\ne 0$. Then $F=0$. 
\item \underline{\bf Case 2}: Let $\Delta_1'=\Delta_2'=0$. Then $F\ne 0$ is possible. 
\end{enumerate}
}
Before we enter into this distinction, 
we write down the conditions for the last two-point function $H=\langle \psi_1\psi_2\rangle$. 
Proceeding as in section~3.1, we find
\BEA
\left( -t \partial_t-\vec{x}\cdot\vec{\partial}-\Delta_1-\Delta_2\right) H 
- \Delta_1' G_{12} -\Delta_2' G_{21} &=& 0 \nonumber \\
\left( -t\vec{\partial} - \vec{\gamma}_1 -\vec{\gamma}_2 - \vec{\chi}_1 -\vec{\chi}_2 \right) H 
- \vec{\gamma}_1' G_{12} -\vec{\gamma}_2'G_{21} &=& 0 
\nonumber \\
\left( -t\left(\vec{\gamma}_1-\vec{\gamma}_2 +\vec{\chi}_1 -\vec{\chi}_2\right) -2\xi_1 \weps \vec{r} \right) H 
-t\left( \vec{\gamma}_1' G_{12} - \vec{\gamma}_2'G_{21}\right) &=& 0
\label{A4} \\
\left( -t\vec{r}\cdot\vec{\partial} -t(\Delta_1-\Delta_2) 
-2\vec{r}\cdot\left(\vec{\chi}_1+\vec{\gamma}_1\right)\right)H & & 
\nonumber \\
-t\left( \Delta_1' G_{12} -\Delta_2' G_{21}\right) -2\vec{\gamma}_1'\cdot\vec{x} G_{12} &=& 0 
\nonumber 
\EEA
Again, we multiply the 3$^{\rm rd}$ of these by $\vec{x}$ and re-use 
$\vec{Y}_0$-covariance on the 4$^{\rm th}$, along with the
known constraint $\Delta_1=\Delta_2$. This gives simultaneously
\BEA
\vec{x} \cdot \left( \vec{\chi}_1-\vec{\chi}_2 +\vec{\gamma}_1 -\vec{\gamma}_2 \right) H 
+  \vec{x}\cdot\left(\vec{\gamma}_1'G_{12} -\vec{\gamma}_2' G_{21}\right) &=& 0
\\
\vec{x} \cdot \left( \vec{\chi}_1-\vec{\chi}_2 +\vec{\gamma}_1 -\vec{\gamma}_2 \right) H 
+  \vec{x}\cdot\left(\vec{\gamma}_1'G_{12} -\vec{\gamma}_2' G_{21}\right)
+  t\left(\Delta_1' G_{12} - \Delta_2' G_{21}\right) &=& 0
\nonumber
\EEA
which implies the constraint
\BEQ
t\left( \Delta_1' G_{12} - \Delta_2' G_{21}\right) = 0
\EEQ
In case 2, we have $\Delta_1'=\Delta_2'=0$ and this constraint is already satisfied. 
In case 1, $F=0$ and the form of $G_{12}$ and $G_{21}$
can be read off from the non-logarithmic representation of section~3.1. 
Since under the exchange $1\leftrightarrow 2$ one has
$G_{12}\leftrightarrow G_{21}$, one can always arrange their amplitudes such that
\BEA
\Delta_1' &=& \Delta_2' \\
G(t,\vec{x}) = G_{12} &=& G_{21} = G(-t,-\vec{x}) 
\nonumber
\EEA
Now, we can consider the two cases separately and work out the two-point functions explicitly.\\

\noindent
\underline{\bf Case 1}. With 
the two constraints $\Delta_1=\Delta_2$ and $\Delta_1'=\Delta_2'$, 
$H$ is to be found from the three independent equations
\BEA
\left( t \partial_t+\vec{x}\cdot\vec{\partial}+2\Delta_1\right) H + 2\Delta_1' G &=& 0 \nonumber \\
\left( t\vec{\partial} + \vec{\gamma}_1 +\vec{\gamma}_2 + \vec{\chi}_1 +\vec{\chi}_2 \right) H 
+ \left(\vec{\gamma}_1' +\vec{\gamma}_2'\right) G &=& 0 
\label{A8} \\
\left( \left(\vec{\gamma}_1-\vec{\gamma}_2 +\vec{\chi}_1 -\vec{\chi}_2\right) +2\xi_1 \weps \vec{u} \right) H 
+\left( \vec{\gamma}_1' - \vec{\gamma}_2'\right) G &=& 0
\nonumber
\EEA
We also require the explicit form of the operators $\vec{\chi}_i$
\BEQ
\left( \vec{\chi}_1+\vec{\chi}_2\right) f 
= \left( \partial_{\vec{\nu}^+} -\xi_1 \weps\vec{\nu}^- \right)f \;\; , \;\;
\left( \vec{\chi}_1-\vec{\chi}_2\right) f = \left( \partial_{\vec{\nu}^-} -\xi_1 \weps\vec{\nu}^+ \right)f
\EEQ
with the variables $\vec{\nu}^{\pm} := \demi \left( \vec{\nu}_1\pm\vec{\nu}_2\right)$. 
Since $F=0$, the mixed correlator $G$
can be read off from (\ref{A1}). The result has already been obtained in section~3.1 and reads
\BEA
G &=& |t|^{-2\Delta_1}\, e^{-(\vec{\gamma}_1+\vec{\gamma}_2)\cdot\vec{u}} g(\vec{u},\vec{\nu}^+,\vec{\nu}^-)
\nonumber \\
&=& g_0(\vec{w}) |t|^{-2\Delta_1}\, e^{-(\vec{\gamma}_1+\vec{\gamma}_2)\cdot\vec{u} 
-(\vec{\gamma}_1-\vec{\gamma}_2)\cdot\vec{\nu}^-}
e^{\xi_1 (\vec{u}+\vec{w})\weps\vec{\nu}^-}
\EEA
where $\vec{w}:=\vec{u}-\vec{\nu}^+$ and $g_0(\vec{w})$ is an undetermined function. Analogously, we write
\BEQ
H = |t|^{-2\Delta_1}\, e^{-(\vec{\gamma}_1+\vec{\gamma}_2)\cdot\vec{u}} h(t,\vec{u},\vec{\nu}^+,\vec{\nu}^-)
\EEQ
and proceed to extract $h$ from the three conditions (\ref{A8}). 
The first one reduces to $t\partial_t h +2\Delta_1'g=0$ and has the solution
\BEQ
h(t,\vec{u},\vec{\nu}^+,\vec{\nu}^-) 
= -2\Delta_1' \ln |t| \: g(\vec{u},\vec{\nu}^+,\vec{\nu}^-) +h_1(\vec{u},\vec{\nu}^+,\vec{\nu}^-)
\EEQ
Next, the second condition (\ref{A8}) becomes
\BEQ
\left( \partial_{\vec{u}} + \partial_{\vec{\nu}^+} 
-\xi_1 \weps \vec{\nu}^- \right) h_1 +\left( \vec{\gamma}_1' +\vec{\gamma}_2'\right) g=0
\EEQ
This is solved by the transformation $h_1 =e^{\xi_1 \vec{u}\weps\vec{\nu}^-} \mathfrak{h}_1$ 
such that the resulting equation for
$\mathfrak{h}_1$ is readily integrated,  with the result
\BEQ \label{A14}
h_1 = -\vec{u}\cdot\left(\vec{\gamma}_1'+\vec{\gamma}_2'\right) g_0(\vec{w}) 
e^{-(\vec{\gamma}_1-\vec{\gamma}_2)\cdot\vec{\nu}^-}
e^{\xi_1 \vec{u}\weps\vec{\nu}^-} + e^{\xi_1 \vec{u}\weps\vec{\nu}^-} h_2(\vec{w},\vec{\nu}^-)
\EEQ
The last condition (\ref{A8}) has the form
\BEQ
\left(\partial_{\vec{\nu}^-} +\xi_1 \weps (\vec{u}+\vec{w}) +\vec{\gamma}_1 - \vec{\gamma}_2 \right)h_1 + 
\left( \vec{\gamma}_1' - \vec{\gamma}_2' \right) g =0
\EEQ
Inserting $h_1$ from (\ref{A14}), and with the transformation $h_2 = e^{-\vec{\nu}^-\weps\vec{w}} 
e^{-\vec{\nu}^-\cdot(\vec{\gamma}_1'-\vec{\gamma}_2')} \mathfrak{h}_2(\vec{w},\vec{\nu}^-)$, this reduces to
$\partial_{\vec{\nu}^-}\mathfrak{h}_2 +(\vec{\gamma}_1'-\vec{\gamma}_2')g_0(\vec{w})$. Hence
$\mathfrak{h}_2 = -\vec{\nu}^-\cdot(\vec{\gamma}_1'-\vec{\gamma}_2') g_0(\vec{w}) + h_0(\vec{w})$ 
such that finally
\BEQ
h_1 = -\vec{u}\cdot\left(\vec{\gamma}_1'+\vec{\gamma}_2'\right) g_0(\vec{w}) 
e^{-(\vec{\gamma}_1-\vec{\gamma}_2)\cdot\vec{\nu}^-}
e^{\xi_1 (\vec{u}+\vec{w})\weps\vec{\nu}^-} 
+ h_0(\vec{w}) e^{-(\vec{\gamma}_1-\vec{\gamma}_2)\cdot\vec{\nu}^-} e^{\xi_1 (\vec{u}+\vec{w})\weps\vec{\nu}^-}
\EEQ
and where $h_0(\vec{w})$ remains undetermined. Summarising, we have found that $F=0$ and 
\BEA
G &=& |t|^{-2\Delta_1} e^{-(\vec{\gamma}_1+\vec{\gamma}_2)\cdot\vec{u} 
-(\vec{\gamma}_1-\vec{\gamma}_2)\cdot\vec{\nu}^-}
e^{\xi_1 (\vec{u}+\vec{w})\wedge\vec{\nu}^-} g_0(\vec{w}) 
\nonumber \\
H &=& |t|^{-2\Delta_1} e^{-(\vec{\gamma}_1+\vec{\gamma}_2)\cdot\vec{u} 
-(\vec{\gamma}_1-\vec{\gamma}_2)\cdot\vec{\nu}^-}
e^{\xi_1 (\vec{u}+\vec{w})\wedge\vec{\nu}^-} h(\vec{u},\vec{w},\vec{\nu}^-) 
\label{A17} \\
h &=& h_0(\vec{w}) - g_0(\vec{w}) \left( 2\Delta_1' \ln|t| + \vec{u}\cdot\left(\vec{\gamma}_1'+\vec{\gamma}_2'\right)
+\vec{\nu}^-\cdot\left(\vec{\gamma}_1'-\vec{\gamma}_2'\right) \right)
\nonumber
\EEA
We have the constraints $\Delta_1=\Delta_2$, $\Delta_1'=\Delta_2'$ and
$\xi_1+\xi_2=0$. At this stage, the functions $g_0(\vec{w})$ and $h_0(\vec{w})$ 
remain undetermined. 

The further consequences of rotation-invariance are discussed in appendix~B. 

\noindent
\underline{\bf Case 2}. Since now $\Delta_1'=\Delta_2'=0$, the first mixed correlator $G_{12}$ 
is obtained from the first three equations (\ref{A1}). 
Similarly, the other mixed correlator $G_{21}$ is found from the equations
\BEA
\left( t \partial_t+\vec{x}\cdot\vec{\partial}+2\Delta_1\right) G_{21} &=& 0 \nonumber \\
\left( t\vec{\partial} + \vec{\gamma}_1 +\vec{\gamma}_2 
+ \vec{\chi}_1 +\vec{\chi}_2 \right) G_{21} + \vec{\gamma}_1'F &=& 0 
\nonumber \\
\left( \left(\vec{\gamma}_1-\vec{\gamma}_2 +\vec{\chi}_1 
-\vec{\chi}_2\right) +2\xi_1 \weps \vec{u} \right) G_{12} +\vec{\gamma}_1'F &=& 0
\nonumber 
\EEA
and the last one is determined from 
\BEA
\left( t \partial_t+\vec{x}\cdot\vec{\partial}+2\Delta_1\right) H  &=& 0 \nonumber \\
\left( t\vec{\partial} + \vec{\gamma}_1 +\vec{\gamma}_2 + \vec{\chi}_1 +\vec{\chi}_2 \right) H 
+ \vec{\gamma}_1' G_{12} +\vec{\gamma}_2' G_{21} &=& 0 
\label{A18} \\
\left( \left(\vec{\gamma}_1-\vec{\gamma}_2 +\vec{\chi}_1 -\vec{\chi}_2\right) +2\xi_1 \weps \vec{u} \right) H 
+ \vec{\gamma}_1' G_{12}- \vec{\gamma}_2' G_{21} &=& 0
\nonumber
\EEA
and all subject to the constraints $\Delta_1=\Delta_2$ and $\xi_1+\xi_2=0$. 
Since $F$ was already found in section~3.1, 
we can also adapt eq.~(\ref{A8}) from the Case 1 treated above and write 
directly down the mixed correlators, with the result
\BEA
F &=& |t|^{-2\Delta_1} e^{-(\vec{\gamma}_1+\vec{\gamma}_2)\cdot\vec{u} 
-(\vec{\gamma}_1-\vec{\gamma}_2)\cdot\vec{\nu}^-}
e^{\xi_1 (\vec{u}+\vec{w})\wedge\vec{\nu}^-} f_0(\vec{w}) 
\nonumber \\
G_{12} &=& |t|^{-2\Delta_1} e^{-(\vec{\gamma}_1+\vec{\gamma}_2)\cdot\vec{u} 
-(\vec{\gamma}_1-\vec{\gamma}_2)\cdot\vec{\nu}^-}
e^{\xi_1 (\vec{u}+\vec{w})\wedge\vec{\nu}^-} \left[ g_{0}(\vec{w}) 
-f_0(\vec{w})\left( \vec{u}-\vec{\nu}^-\right)\cdot\vec{\gamma}_2'\right]~~
\label{A19} \\
G_{21} &=& |t|^{-2\Delta_1} e^{-(\vec{\gamma}_1+\vec{\gamma}_2)\cdot\vec{u} 
-(\vec{\gamma}_1-\vec{\gamma}_2)\cdot\vec{\nu}^-}
e^{\xi_1 (\vec{u}+\vec{w})\wedge\vec{\nu}^-} \left[ g_{0}(\vec{w}) 
-f_0(\vec{w})\left( \vec{u}+\vec{\nu}^-\right)\cdot\vec{\gamma}_1'\right]
\nonumber 
\EEA
Herein, we have taken into account that under the exchange $1\leftrightarrow 2$ of the sites, 
one has the permutation $G_{12}\leftrightarrow G_{21}$. Then $f_0$ and $g_0$ remain undetermined. 
The correlator $H$ is written in the form 
\BEQ \label{A20}
H = |t|^{-2\Delta_1} e^{-(\vec{\gamma}_1+\vec{\gamma}_2)\cdot\vec{u} 
-(\vec{\gamma}_1-\vec{\gamma}_2)\cdot\vec{\nu}^-}
e^{\xi_1 (\vec{u}+\vec{w})\wedge\vec{\nu}^-} h(\vec{u},\vec{w},\vec{\nu}^-) 
\EEQ
Then the first of eqs.~(\ref{A18}) is automatically satisfied, whereas the second and  third lead to the system
\BEA
\partial_{\vec{u}} h +\left( \vec{\gamma}_1'+\vec{\gamma}_2'\right) g_0 
- \left( \vec{\gamma}_1'\: \left( \vec{u}-\vec{\nu}^-\right)\cdot\vec{\gamma}_2' 
+ \vec{\gamma}_2'\: \left( \vec{u}+\vec{\nu}^-\right)\cdot\vec{\gamma}_1' \right) f_0 &=& 0 
\nonumber \\
\partial_{\vec{\nu}^-} h +\left( \vec{\gamma}_1'-\vec{\gamma}_2'\right) g_0 
- \left( \vec{\gamma}_1'\: \left( \vec{u}-\vec{\nu}^-\right)\cdot\vec{\gamma}_2' 
- \vec{\gamma}_2'\: \left( \vec{u}+\vec{\nu}^-\right)\cdot\vec{\gamma}_1' \right) f_0 &=& 0 
\EEA
These are decoupled by defining $\vec{y}^{\pm} :=\demi \left( \vec{u}\pm \vec{\nu}^-\right)$. 
Considering $h=h(\vec{y}^+,\vec{y}^-,\vec{w})$, one has
\BEQ
\partial_{\vec{y}^+}h = - 2\vec{\gamma}_1' g_0 +2 \vec{\gamma}_1'\: \vec{y}^-\cdot\vec{\gamma}_2' f_0 \;\; , \;\;
\partial_{\vec{y}^-}h = - 2\vec{\gamma}_2' g_0 +2 \vec{\gamma}_2'\: \vec{y}^+\cdot\vec{\gamma}_1' f_0 
\EEQ
such that finally, with $h_0(\vec{w})$ an undetermined function
\BEA
h      &=& h_0(\vec{w}) -2g_{0}(\vec{w}) \vec{y}^+\cdot\vec{\gamma}_1' 
- 2g_{0}(\vec{w})   \vec{y}^-\cdot\vec{\gamma}_2' 
+ 2f_0(\vec{w}) \vec{y}^+\cdot\vec{\gamma}_1' \: \vec{y}^-\cdot\vec{\gamma}_2' 
\nonumber \\
       &=& h_0(\vec{w}) -g_{0}(\vec{w}) \left( \vec{u}+\vec{\nu}^-\right)\cdot\vec{\gamma}_1' 
- g_{0}(\vec{w}) \left( \vec{u}-\vec{\nu}^-\right)\cdot\vec{\gamma}_2' \nonumber \\
& & +\demi f_0(\vec{w}) \left( \vec{u}+\vec{\nu}^-\right)\cdot\vec{\gamma}_1' \: \left( 
\vec{u}-\vec{\nu}^-\right)\cdot\vec{\gamma}_2'
\label{A23}
\EEA
The discussion of rotation-invariance is analogous to Case 1 and carried out in appendix~B.  
Combining the results eqs.~(\ref{A19},\ref{A20},\ref{A23}) 
and reverting to the original coordinates, one arrives at the
final form (\ref{ecga_case2},\ref{ecga_gh}) stated in the main text.

\appsection{B}{On rotation-invariance in the {\sc ecga}}

Having discussed in the main text the shape of two-point functions co-variant under the {\sc ecga}, we now consider the additional
consequences when rotation-invariance is taken into account as well. 

\subsection{Rotation-invariance for vanishing rapidities} 

We shall compare the consequences of using two distinct representations for the infinitesimal generator of rotations, 
namely (also recall that $\vec{a}\wedge \vec{b} = \epsilon_{ij} a_i b_j$)
\BEA
J &:=& - \epsilon_{ij} x_i \frac{\partial}{\partial x_j} - \frac{1}{2\xi} \chi_i \chi_i 
= - \vec{x}\wedge\partial_{\vec{x}} - \frac{1}{2\xi} \vec{\chi}^2
\nonumber \\
R &:=& -\epsilon_{ij} x_i \frac{\partial}{\partial x_j} - \epsilon_{ij} \nu_i \frac{\partial}{\partial \nu_j} 
= -\vec{x}\wedge \partial_{\vec{x}} - \vec{\nu}\wedge \partial_{\vec{\nu}}
\EEA
Martelli and Tachikawa \cite{Tachikawa} advocated in favour of the generator $J$, because it naturally appears in the contraction
procedure they used in deriving the {\sc ecga}. Here, we wish to compare with the results found for the naturally-looking generator $R$, 
also mentioned as a possible alternative in \cite{Tachikawa}. 

{}From a purely algebraic point of view, there is no criterion which would lead one to prefer one of these two choices. Both
obey the same commutation relations with the other generators of the {\sc ecga} and both commute with the Schr\"odinger operator
$\cal S$ defined in section~3. 

Here, we shall show that physically distinct results are found, depending on the use of $J$ (which we shall refer to as
`{\em $J$-invariance}') or $R$ (which we shall refer to as `{\em $R$-invariance}'). 
Namely, the rapidity-less {\sc ecga}-covariant two-point function
$F=\langle \phi_1 \phi_2\rangle$ has the form
\BEQ \label{B2}
F = f_0(\vec{u}-\vec{\nu}^+) t^{-2\Delta_1} e^{-\xi_1 \vec{\nu}^- \wedge (2\vec{u} -\vec{\nu}^+)}\, 
\delta_{\Delta_1,\Delta_2}\delta_{\xi_1+\xi_2,0}
\EEQ
where $\vec{\nu}^{\pm}=\demi \left(\vec{\nu}_1 \pm \vec{\nu}_2\right)$ and still contains an undetermined differentiable function 
 $f_0=f_0(\vec{w})$ of the single variable $\vec{w}=\vec{u}-\vec{\nu}^+$. 
Any explicit dependence on the single variables $\vec{\nu}^{\pm}$ of the
two-point function $F$ is already contained in (\ref{B2}). 
{\em The additional requirement of rotation-invariance leads to a clear distinction}
\BEQ \label{B3}
\left\{ \begin{array}{ll} \mbox{\rm $f_0$ is arbitrary} & \mbox{\rm ~~;~ $J$-invariance} \\ 
\mbox{\rm $f_0=f_0(w_1^2+w_2^2)$} & \mbox{\rm ~~;~ $R$-invariance}
\end{array} \right.
\EEQ

\noindent 
{\bf Proof:} 
Begin by writing down the two-particle form of the generators $J$ and $R$ (where in view of the coming application to $F$, 
the Bargman super-selection rule $\xi_2 = - \xi_1$ has already been used)
\BEA 
J &=& - \vec{u}\wedge\partial_{\vec{u}} 
-\frac{1}{2\xi_1} \left( \partial_{\vec{\nu}^+}\cdot\partial_{\vec{\nu}^-} +\xi_1^2\, \vec{\nu}^+\cdot\vec{\nu}^-\right) 
-\demi \left( \vec{\nu}^+\wedge\partial_{\vec{\nu}^+} +  \vec{\nu}^-\wedge\partial_{\vec{\nu}^-} \right) 
\nonumber \\
R &=& -\vec{u}\wedge\partial_{\vec{u}} 
- \vec{\nu}_1\wedge\partial_{\vec{\nu}_1}- \vec{\nu}_2\wedge\partial_{\vec{\nu}_2}
\:=\:  -\vec{u}\wedge\partial_{\vec{u}} 
- \vec{\nu}^+\wedge\partial_{\vec{\nu}^+}- \vec{\nu}^-\wedge\partial_{\vec{\nu}^-}
\label{B4}
\EEA 
In working out the condition of rotation-invariance $JF \stackrel{!}{=}0$ or $RF \stackrel{!}{=}0$, respectively, 
we shall need the following auxiliary formul{\ae}, with
{\small$\weps = \left( \matz{0}{1}{-1}{0}\right)$} and $\vec{a},\vec{b}\in\mathbb{R}^2$ 
such that $\vec{a}\wedge\vec{b}=\vec{a}\cdot\left(\weps \vec{b}\right)$
\BEA
\partial_{\vec{a}}\: e^{-\vec{a}\wedge\vec{b}} &=& - \left( \weps \vec{b} \right)\, e^{-\vec{a}\wedge\vec{b}} \nonumber \\
\left( \weps \vec{a}\right) \cdot \vec{b} &=& - \vec{a} \wedge \vec{b}  \label{B5}\\
\left( \weps \vec{a}\right) \cdot \left( \weps \vec{b}\right)  &=&  \vec{a} \cdot \vec{b} \nonumber \\
\vec{a} \wedge \left( \weps \vec{b}\right)  &=&  -\vec{a} \cdot \vec{b} \nonumber 
\EEA
Then, application of the two distinct rotation generators to the two-point function (\ref{B2}) leads straightforwardly 
in the case of
$J$-invariance simply to the identity $JF=0$, whereas in the case of $R$-invariance we obtain
$\vec{w} \wedge \partial_{\vec{w}} f_0(\vec{w})=0$. Hence, given the form (\ref{B2}), $J$-invariance is automatic and 
does not impose any further condition on the function
$f_0$. On the other hand, in the case of $R$-invariance, $f_0$ can only be a function of the scalar $|\vec{w}|^2=w_1^2+w_2^2$. 
This proves the assertion (\ref{B3}). \hfill q.e.d. 

As we shall see below, this distinction between the generators $J$ and $R$ 
will lead to further consequences in the case of non-vanishing rapidities. 

\subsection{Rotation-invariance for non-vanishing rapidities: non-logarithmic case} 

In the non-logarithmic case, the {\sc ecga}-covariant two-point function reads 
\BEQ \label{B6}
F = f_0(\vec{u}-\vec{\nu}^+) t^{-2\Delta_1} e^{-2\vec{\gamma}_+\cdot\vec{u} -2\vec{\gamma}_-\cdot\vec{\nu}^-}\, 
e^{-\xi_1 \vec{\nu}^- \wedge (2\vec{u} -\vec{\nu}^+)}\, 
\delta_{\Delta_1,\Delta_2}\delta_{\xi_1+\xi_2,0}
\EEQ
where in addition to the previous conventions, we also defined 
$\vec{\gamma}_{\pm} := \demi \left( \vec{\gamma}_1\pm\vec{\gamma}_2\right)$. Therefore, the two rotation-generators must be generalised
to include rapidity terms and read for a single particle
\BEQ
J = -\vec{x}\wedge\partial_{\vec{x}} - \vec{\gamma} \wedge \partial_{\vec{\gamma}} - \frac{1}{2\xi} \vec{\chi}\cdot\vec{\chi}
\;\; , \;\;
R = -\vec{x}\wedge\partial_{\vec{x}} - \vec{\gamma} \wedge \partial_{\vec{\gamma}} - \vec{\nu}\wedge \partial_{\vec{\nu}}
\EEQ
and for two particles
\BEA 
J &=& - \vec{u}\wedge\partial_{\vec{u}} 
-\frac{1}{2\xi_1} \left( \partial_{\vec{\nu}^+}\cdot\partial_{\vec{\nu}^-} +\xi_1^2\, \vec{\nu}^+\cdot\vec{\nu}^-\right) 
-\demi \left( \vec{\nu}^+\wedge\partial_{\vec{\nu}^+} +  \vec{\nu}^-\wedge\partial_{\vec{\nu}^-} \right) 
\nonumber \\
 & & -\vec{\gamma}_+\wedge\partial_{\vec{\gamma}_+} -\vec{\gamma}_-\wedge\partial_{\vec{\gamma}_-} 
 \nonumber \\
R &=&  -\vec{u}\wedge\partial_{\vec{u}} - \vec{\gamma}_+\wedge\partial_{\vec{\gamma}_+}- \vec{\gamma}_-\wedge\partial_{\vec{\gamma}_-} 
- \vec{\nu}^+\wedge\partial_{\vec{\nu}^+}- \vec{\nu}^-\wedge\partial_{\vec{\nu}^-}
\label{B8}
\EEA 
Again, the undetermined function $f_0=f_0(\vec{\gamma}_+,\vec{\gamma}_-,\vec{w})$ 
depends on the single variable $\vec{w}=\vec{u}-\vec{\nu}^+$. However, since the rotations can also transform the rapidities
$\vec{\gamma}_{\pm}$, $f_0$ can in addition also depend explicitly on them. Hence, $f_0=f_0(\vec{\gamma}_+,\vec{\gamma}_-,\vec{w})$ 
will be a function of 6 variables, subject to
a single condition coming form rotation-invariance. 

Using the same auxiliary identities (\ref{B5}) as before, straightforward but slightly tedious calculations lead to
\BEA
\mbox{\rm $J$-invariance} &:& 
\vec{\gamma}_+\wedge\frac{\partial f_0}{\partial\vec{\gamma}_+} + \vec{\gamma}_-\wedge\frac{\partial f_0}{\partial\vec{\gamma}_-}
+\frac{\vec{\gamma}_-}{\xi_1} \cdot \frac{\partial f_0}{\partial \vec{w}} = 0 
\nonumber \\
\mbox{\rm $R$-invariance} &:& 
\vec{\gamma}_+\wedge\frac{\partial f_0}{\partial\vec{\gamma}_+} + \vec{\gamma}_-\wedge\frac{\partial f_0}{\partial\vec{\gamma}_-}
\label{B9}
+ \vec{w} \wedge \frac{\partial f_0}{\partial \vec{w}} = 0
\EEA
In order to find the general solutions of these, in the case of $J$-invariance, one introduces a new variable
$\vec{v} := \vec{\gamma}_--\xi_1 \weps \vec{w}$ and takes $f_0$ as a function $f_0(\vec{\gamma}_+,\vec{\gamma}_-,\vec{v})$.  
Then the first of eqs.~(\ref{B9}) reduces to 
$\left(\vec{\gamma}_+\wedge \partial_{\vec{\gamma}_+} + \vec{\gamma}_-\wedge\partial_{\vec{\gamma}_-}\right) f_0=0$. 
Three obvious and independent solutions of this are $\vec{\gamma}_+^2$, $\vec{\gamma}_-^2$ and $\vec{\gamma}_+\cdot\vec{\gamma}_-$, 
from which the general solution can be constructed. On the other hand, in the case of $R$-invariance one easily lists 5 independent
solutions so that finally
\BEA
\mbox{\rm $J$-invariance} &:& f_0=f_0\left(\vec{\gamma}_+^2, \vec{\gamma}_-^2, \vec{\gamma}_+\cdot\vec{\gamma}_-, 
\vec{w}+\weps \vec{\gamma}_- \xi_1^{-1} \right) 
\nonumber \\
\mbox{\rm $R$-invariance} &:& f_0=f_0\left(\vec{w}^2,\vec{\gamma}_+^2, \vec{\gamma}_-^2, \vec{w}\cdot\vec{\gamma}_+,
\vec{w}\cdot\vec{\gamma}_-\right)
\EEA
Reverting to the original variables $\vec{\gamma}_{1,2}$ gives the expressions in the main text or appendix~A. 

A comparison of the two distinct eqs.~(\ref{B9}) shows the origin of these two distinct forms of the function $f_0$: while in the
case of $R$-invariance, the habitual form the of the rotation generator guarantees that formal scalar products of the vectors
$\vec{w}$, $\vec{\gamma}_+$ and $\vec{\gamma}_-$ are always rotation-invariant, this holds no longer true in the case of $J$-invariance, 
where only scalar products between the rapidity vectors $\vec{\gamma}_{\pm}$ have this property. Model-specific calculations will
permit to distinguish between these possibilities. 

We also observe that in the case of $J$-invariance, taking the non-exotic limit $\xi_1\to 0$ enforces $\vec{\gamma}_-\to \vec{0}$, 
in order to maintain a finite value in the last argument of the function $f_0$. In this way, one can understand how the
constraint $\vec{\gamma}_1=\vec{\gamma}_2$ in non-exotic CGA is recovered. 
No such limit argument can be made in the case of $R$-invariance.  

\subsection{Rotation-invariance for non-vanishing rapidities: logarithmic case} 

One must must take further into account that the $\vec{\gamma}$'s become Jordan matrices, 
such that the rotation generators have to be
generalised to the forms
\BEQ \label{B11}
J = -\vec{x}\wedge\partial_{\vec{x}} - \vec{\gamma} \wedge \partial_{\vec{\gamma}} - \vec{\gamma}' \wedge \partial_{\vec{\gamma}'}
- \frac{1}{2\xi} \vec{\chi}\cdot\vec{\chi}
\;\; , \;\;
R = -\vec{x}\wedge\partial_{\vec{x}} - \vec{\gamma} \wedge \partial_{\vec{\gamma}} - \vec{\gamma}' \wedge \partial_{\vec{\gamma}'}
- \vec{\nu}\wedge \partial_{\vec{\nu}}
\EEQ
for a single particle and with analogous extensions in the two-particle case. In addition to the two-point
function $F=\langle \phi_1 \phi_2\rangle$ already analysed in the non-logarithmic representations, one now has to consider
the additional two-point functions $G=\langle \phi_1 \psi_2\rangle$ and $H=\langle \psi_1\psi_2\rangle$. Furthermore, we have already
seen in the main text that two distinct cases have to be distinguished, depending on whether the matrices $\wht{\Delta}_{1,2}$ 
have Jordan form (case 1) or are diagonal (case 2). 

{\bf A)} If we consider $R$-invariance, begin with case 1 (where $F=0$ and require co-variance $RG \stackrel{!}{=} 0 \stackrel{!}{=} RH$, one has
\BEQ 
\vec{w}\wedge \frac{\partial g_0}{\partial \vec{w}} 
+ \vec{\gamma}_1\wedge \frac{\partial g_0}{\partial \vec{\gamma}_1} 
+ \vec{\gamma}_2\wedge \frac{\partial g_0}{\partial \vec{\gamma}_2}  
+ \vec{\gamma}_1'\wedge \frac{\partial g_0}{\partial \vec{\gamma}_1'} 
+ \vec{\gamma}_2'\wedge \frac{\partial g_0}{\partial \vec{\gamma}_2'} =0 
\EEQ
and analogously for $h_0$, since all individual terms in (\ref{A17}) are explicitly rotation-invariant. In principle, the functions
$g_0$, $h_0$ depend on the 10 variables $\vec{w}$, $\vec{\gamma}_{1,2}$, $\vec{\gamma}_{1,2}'$. Since rotation-invariance imposes a
single extra condition, there remains a function of 9 rotation-invariant variables, for example
\BEA 
g_0=g_0\left( \vec{w}^2,\vec{\gamma}_1^2,\vec{\gamma}_2^2,{\vec{\gamma}_1'}^2, 
{\vec{\gamma}_2'}^2,\vec{w}\cdot\vec{\gamma}_1,\vec{w}\cdot\vec{\gamma}_2,
\vec{w}\cdot\vec{\gamma}_1',\vec{w}\cdot\vec{\gamma}_2'\right)
\nonumber \\
h_0=h_0\left( \vec{w}^2,\vec{\gamma}_1^2,\vec{\gamma}_2^2,{\vec{\gamma}_1'}^2, 
{\vec{\gamma}_2'}^2,\vec{w}\cdot\vec{\gamma}_1,\vec{w}\cdot\vec{\gamma}_2,
\vec{w}\cdot\vec{\gamma}_1',\vec{w}\cdot\vec{\gamma}_2'\right)
\label{B13}
\EEA
For case 2, an analogous argument applies and one recovers (\ref{B13}) along with an analogous form for $f_0$. 

{\bf B)} A very different result is found for $J$-invariance. If one considers case 1 first, one has again $F=\langle \phi_1 \phi_2\rangle =0$,
whereas $G=\langle \phi_1 \psi_2\rangle$ has the same form as the two-point function $F$ treated above in the non-logarithmic case. 
It remains to consider the two-point function
\BEQ
H = \langle \psi_1 \psi_2\rangle = t^{-2\Delta_1} e^{-2\vec{\gamma}_+\cdot\vec{u} -2\vec{\gamma}_-\cdot\vec{\nu}^-}\, h
\EEQ
where the scaling function $h$ can be written as
\BEQ
h = h_0(\vec{u}-\vec{\nu}^+) - g_0(\vec{u}-\vec{\nu}^+) \left( 2\Delta_1'\ln |t| + \vec{u}\cdot\left(\vec{\gamma}_1'+\vec{\gamma}_2'\right) 
+ \vec{\nu}^{-}\cdot\left(\vec{\gamma}_1'-\vec{\gamma}_2'\right) \right)
\EEQ
In complete analogy to the previous sub-section, $J$-invariance implies the conditions
\BEQ
{\cal D} h = 0 \;\; , \;\; {\cal D} g_0 = 0
\EEQ
with the differential operator
\BEQ
{\cal D} := 
  \vec{\gamma}_+\wedge \frac{\partial }{\partial \vec{\gamma}_+} 
+ \vec{\gamma}_-\wedge \frac{\partial }{\partial \vec{\gamma}_-}  
+ \vec{\gamma}_1'\wedge \frac{\partial }{\partial \vec{\gamma}_1'} 
+ \vec{\gamma}_2'\wedge \frac{\partial }{\partial \vec{\gamma}_2'} 
+\frac{\vec{\gamma}_-}{\xi_1} \cdot \frac{\partial }{\partial \vec{w}}
\EEQ
This gives the following equation for $h_0$: 
\BEQ
{\cal D} h_0 - g_0 {\cal D} \left( 2\Delta_1'\ln |t| + \vec{u}\cdot\left(\vec{\gamma}_1'+\vec{\gamma}_2'\right) 
+ \vec{\nu}^{-}\cdot\left(\vec{\gamma}_1'-\vec{\gamma}_2'\right)\right) =0 
\EEQ
Working out the differential operator and taking the condition $\vec{w}=\vec{u}-\vec{\nu}^+$ into account, leads to
\BEQ
{\cal D}h_0 - \left( \left(\vec{\gamma}_1'+\vec{\gamma}_2'\right) \cdot \left( \frac{1}{\xi_1} \vec{\gamma}_- +\vec{u} \right)
+ \left(\vec{\gamma}_1'-\vec{\gamma}_2'\right)\wedge \vec{\nu}^- \right) g_0 =0
\EEQ
However, since $h_0$ depends only on $\vec{w}$ and not on $\vec{u}$ or $\vec{\nu}^{\pm}$ separately, this condition is only compatible
with our previous results if
\BEQ 
\vec{\gamma}_1' = \vec{\gamma}_2' = \vec{0}
\EEQ
Then only the matrix $\wht{\Delta}$ of the conformal weight can have a Jordan form and  
\BEA 
g_0=g_0\left( \vec{\gamma}_+^2,\vec{\gamma}_-^2,\vec{\gamma}_+\cdot\vec{\gamma}_-, \vec{w}+\weps \vec{\gamma}_- \xi_1^{-1} \right)
\nonumber \\
h_0=h_0\left( \vec{\gamma}_+^2,\vec{\gamma}_-^2,\vec{\gamma}_+\cdot\vec{\gamma}_-, \vec{w}+\weps \vec{\gamma}_- \xi_1^{-1} \right)
\EEA
Reverting to the original $\vec{\gamma}_{1,2}$ gives the expressions in the main text or in appendix~A. 

Similar arguments apply to case 2:  since now $F=\langle \phi_1\phi_2\rangle\ne 0$, consideration of $G_{12}$ leads to $\vec{\gamma}_2'=\vec{0}$
and the other mixed two-point function $G_{21}$ gives $\vec{\gamma}_1'=\vec{0}$. Then no logarithmic structure remains. 
For $J$-invariance, there is but a single case, see table~\ref{tab1}.


\end{document}